\documentclass[structabstract]{aa}  

\usepackage[colorlinks=true, linkcolor=blue, citecolor=blue, urlcolor=blue]{hyperref}
\usepackage{amsmath}
\usepackage{graphicx}
\usepackage{txfonts}
\usepackage{amssymb}
\usepackage[]{natbib}
\usepackage{mathrsfs}
\usepackage{caption}
\usepackage{float}
\usepackage{afterpage}

\newcommand{\wasp}[0]{\mbox{WASP-33}}
\newcommand{\waspb}[0]{\mbox{WASP-33b}}
\newcommand{\waspa}[0]{\mbox{WASP-33A}}

\newcommand{\bjdtdb}[0]{\mbox{$\mathrm{BJD}_{\mathrm{TDB}}$}}

\begin{document}

   \title{Pulsation analysis and its impact on primary transit
     modeling in \wasp}
   \subtitle{}

   \author{C. von Essen$^{1}$, S. Czesla$^{1}$, U. Wolter$^{1}$, M. Breger$^{2,6}$, E. Herrero$^{3}$, M. Mallonn$^{4}$, I. Ribas$^{3}$, K. G. Strassmeier$^{4}$ and J. C. Morales$^{3,5}$}
   \authorrunning{C. von Essen et al.}
   \titlerunning{Pulsations on WASP-33b}
   \offprints{cessen@hs.uni-hamburg.de}

   \institute{$^1$Hamburger Sternwarte, University of Hamburg,
              Gojenbergsweg 112, 21029 Hamburg, Germany\\
              $^2$Department of Astronomy, University of Texas, Austin, TX 78712, USA\\
              $^3$Institut de Ci\`encies de l’Espai (CSIC-IEEC), Campus UAB, Facultat de Ci\`encies, 
              Torre C5 parell, 2a pl, 08193 Bellaterra, Spain\\
              $^4$Leibniz-Institut f\"ur Astrophysik Potsdam, An der Sternwarte 16, 14482, Potsdam, Germany\\
              $^5$Dept. d’Astronomia i Meteorologia, Institut de Ci\`encies del Cosmos (ICC), 
              Universitat de Barcelona (IEEC-UB), Mart\'{\i} Franqu\`es 1, E08028 Barcelona, Spain\\
              $^6$Institut f\"ur Astronphysik der Universit\"at Wien, T\"urkenschanzstr. 17, A–1180, Wien, Austria\\
              \email{cessen@hs.uni-hamburg.de}
               }

   \date{Received 6/08/2013; accepted 19/10/2013}

\abstract {} 
  {To date, \wasp \ is the only $\delta$ Scuti star known to be
    orbited by a hot Jupiter. The pronounced stellar pulsations,
    showing periods comparable to the primary transit duration,
    interfere with the transit modeling. Therefore our main goal is
    to study the pulsation spectrum of the host star to redetermine
    the orbital parameters of the system by means of pulsation-cleaned
    primary transit light curves.}
  {Between August 2010 and October 2012 we obtained 457 hours of
    photometry of \wasp\ using small and middle-class telescopes
    located mostly in Spain and in Germany. Our observations comprise
    the wavelength range between the blue and the red, and provide
    full phase coverage of the planetary orbit. After a careful
    detrend, we focus our pulsation studies in the high frequency
    regime, where the pulsations that mostly deform the primary
    transit exist.}
  {The data allow us to identify, for the first time in the system,
    eight significant pulsation frequencies. The pulsations are likely
    associated with low-order p-modes. Furthermore, we find that
    pulsation phases evolve in time. We use our knowledge of the
    pulsations to clean the primary transit light curves and carry out
    an improved transit modeling. Surprisingly, taking into
      account the pulsations in the modeling has little influence on
      the derived orbital parameters. However, the uncertainties in
      the best-fit parameters decrease. Additionally,
    we find indications for a possible dependence between wavelength
    and transit depth, but only with marginal significance. A clear
    pulsation solution, in combination with an accurate orbital
    period, allows us to extend our studies and search for star-planet
    interactions (SPI). Although we find no conclusive evidence of
    SPI, we believe that the pulsation nature of the host star and the
    proximity between members make \wasp\ a promising system for
    further SPI studies.}
  {}
          
   \maketitle

\section{Introduction}

The first mention of $\delta$ Scuti variability was made more than one
hundred years ago \citep{Campbell1900}. Fifty years later,
\cite{Eggen1956} pointed out the need to place these variable stars
under an independent stellar classification. The $\delta$ Scutis have
been among us for a long time. Nevertheless, because of the
intrinsically small variability near the limit of detectability major
studies of them did not begin until the 1970s
\citep{Baglin1973,Breger1979,Breger1983}. Nowadays, the Kepler
telescope alone provides precise light curves of several hundred
$\delta$ Scuti stars \citep[e.g.,][]{Uytterhoeven2011}.


In the Hertzsprung-Russell diagram, the $\delta$~Scuti stars are
located in an instability strip covering spectral types between A and
F \citep{Baglin1973,Breger1983}. Most $\delta$ Scuti stars belong to
Population I \citep{Breger1979}, with typical masses of 2~$M_{\odot}$
\citep{Milligan1992}. Due to both radial and nonradial pulsations,
$\delta$~Scuti stars show brightness variations from milli-magnitudes
up to almost one magnitude in blue bands. In $\delta$~Scuti stars,
pulsations are driven by opacity variations. There are two distinct
types of pulsation modes that might occur \citep[e.g.,][]{Breger2012}:
short-period p-modes (pressure modes, for which pressure serves
  as the restoring force) and long-period g-modes (gravity modes,
  with buoyancy as the restoring force). A typical $\delta$~Scuti
pulsation spectrum shows dozens of periods
\citep{Breger1999a,Breger1999b}, with cycle durations ranging from a
couple of hours to the minute regime.

\wasp\ (HD 15082) is a bright (\mbox{$V\sim8.3$}), rapidly rotating
\mbox{($v\sin(i)\sim90$ km/s)} \footnote{$i$ corresponds to the
  inclination of the stellar rotation axis.} $\delta$~Scuti star; in
fact, it is both the hottest and only $\delta$~Scuti star known to
date to host a hot Jupiter \citep{Christian2006}.  The planet, \waspb,
was detected through its transits in the frame of the WASP
campaign \citep{Pollaco2006}.  It circles its host star every $1.22$~d
in a retrograde orbit. With a brightness temperature of 3620~K,
\waspb\ is the hottest exoplanet known to date \citep{Smith2011}.
Showing an unusually large radius, \waspb\ belongs to the class of
anomalously inflated exoplanets \citep[][]{CollierCameron2010}.  For
its mass and hence density, only an upper limit of
\mbox{$M\sin(i)<4.59~M_J$} \footnote{$i$ corresponds to the
  inclination of the planetary orbit.} has been determined.

The host star, \waspa, shows pronounced pulsations with periods on the
order of one hour.  \citet{CollierCameron2010} note that the presence
of these pulsations offers ``the intriguing possibility that tides
raised by the close-in planet may excite or amplify the pulsations in
such stars''. The discovery of \wasp's pulsations within photometric
data were first reported by \cite{Herrero2011}, who suggest a
possible commensurability between a pulsation period and the planetary
orbital period with a factor of 26, indicative of
SPI.

We study the pulsations and primary transits using a total of 56 light
curves of \wasp, observed during two years and providing complete
orbital phase coverage.


\section{Observations and Data Reduction}
\label{Sec2}
\begin{figure}[ht!]
  \centering
  \includegraphics[width=.5\textwidth]{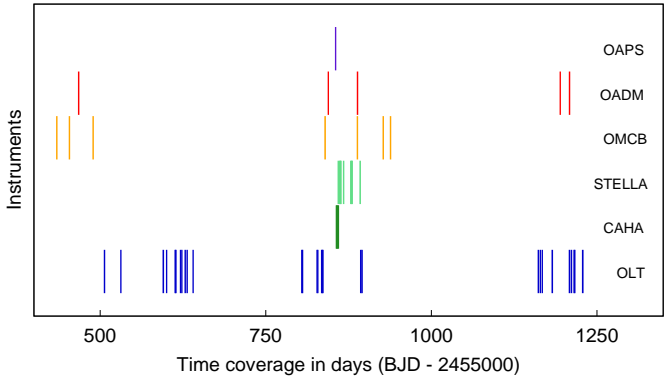}
  \vspace{-1cm}
  \caption{\label{fig:instruments}Sampling of our observations.}
\end{figure}

Between Aug. 2010 and Oct. 2012, we obtained 457 hours of photometry
of \wasp\ distributed across 56 nights using six telescopes: one in
Germany and five in Spain.  Figure~\ref{fig:instruments} shows the
temporal coverage provided by the individual telescopes; the details
of the observations are given in Tables~\ref{tab:OBS_OLT}
and~\ref{tab:OBS_CAHA} and the technical characteristics of the
telescopes are summarized in Table~\ref{tab:telescopeData}.

Throughout the analysis, the barycentric dynamical time system is used
(\bjdtdb).  Conversions between different time reference systems have
been carried out using the web-tool made available by
\citet{Eastman2010}\footnote{\url{http://astroutils.astronomy.ohio-state.edu/time/}}.

WASP-33 is located in a sparse stellar field. Therefore the defocusing
technique did not produce any undesired effect, such as overlapping of
the stellar point spread functions. However, after either defocusing
or considering the natural seeing of the sites, the optical companion
identified by \cite{Moya2011} is contained, in most of the cases,
inside the selected aperture radius. A discussion on third-light
contribution is addressed in Section~\ref{Sec4}.

\begin{table}[h] 
  \caption{\label{tab:OBS_OLT}
    Overview of observation nights for OLT and STELLA.
    }
        \centering
        \begin{tabular}{l r r r r}
          \hline
          \hline
          Date               & Duration & F$^a$  & NoP$^b$     & Airmass \\
          & (h) & & & \\ 
          \hline
          \multicolumn{5}{c}{OLT} \\ \hline
          2010 November 6    & 5.9  & R  & 803     & 1.0$\rightarrow$1.2 \\
          2010 December 1    & 1.4  & B  & 60      & 1.5$\rightarrow$2.1 \\
          2011 February 3    & 5.2  & R  & 869     & 1.0$\rightarrow$1.4 \\
          2011 February 8    & 4.0  & R  & 89      & 1.0$\rightarrow$1.2 \\
          2011 February 21   & 5.9  & R  & 1277    & 1.0$\rightarrow$2.3 \\
          2011 February 22   & 5.3  & R  & 769     & 1.0$\rightarrow$2.3 \\
          2011 March 1       & 4.0  & B  & 83      & 1.0$\rightarrow$1.6 \\
          2011 March 3       & 3.6  & B  & 176     & 1.0$\rightarrow$1.6 \\
          2011 March 8       & 2.5  & B  & 186     & 1.2$\rightarrow$2.3 \\
          2011 March 11      & 4.6  & B  & 296     & 1.1$\rightarrow$2.4 \\
          2011 March 20      & 3.3  & B  & 277     & 1.2$\rightarrow$2.1 \\
          2011 August 31     & 5.5  & B  & 541     & 1.0$\rightarrow$2.4 \\
          2011 September 1   & 5.2  & B  & 906     & 1.0$\rightarrow$2.5 \\
          2011 September 23  & 4.1  & B  & 632     & 1.0$\rightarrow$2.9 \\
          2011 September 24  & 4.3  & B  & 632     & 1.0$\rightarrow$1.2 \\
          2011 September 30  & 5.8  & B  & 754     & 1.1$\rightarrow$2.0 \\
          2011 October 1     & 3.2  & B  & 443     & 1.4$\rightarrow$3.1 \\
          2011 October 2     & 3.5  & B  & 439     & 1.5$\rightarrow$2.2 \\
          2011 November 28   & 7.1  & B  & 650     & 1.0$\rightarrow$1.4 \\
          2011 November 30   & 7.6  & B  & 1174    & 1.0$\rightarrow$2.4 \\
          2012 August 23     & 5.2  & B  & 545     & 1.0$\rightarrow$2.3 \\
          2012 August 25     & 6.5  & B  & 242     & 1.0$\rightarrow$1.8 \\
          2012 August 28     & 3.9  & B  & 243     & 1.3$\rightarrow$2.8 \\
          2012 September 12  & 7.3  & B  & 621     & 1.0$\rightarrow$2.1 \\
          2012 October 8     & 6.6  & B  & 758     & 1.0$\rightarrow$2.4 \\
          2012 October 11    & 8.6  & B  & 1160    & 1.0$\rightarrow$2.0 \\
          2012 October 15    & 2.9  & B  & 241     & 1.3$\rightarrow$2.1 \\
          2012 October 16    & 4.9  & B  & 667     & 1.0$\rightarrow$1.5 \\
          2012 October 28    & 11.8 & B  & 736     & 1.0$\rightarrow$2.6 \\
          \hline
          \multicolumn{5}{c}{STELLA} \\
          \hline
          2011 October 25        & 11    & v    & 78   & 1.0$\rightarrow$2.9 \\
          & 11    & b    & 81   &  \\
          2011 October 27        & 10.4  & v    & 301  & 1.0$\rightarrow$2.6 \\
          & 10.4  & b    & 144  & \\
          2011 October 28        & 10.1  & v    & 322  & 1.0$\rightarrow$2.5 \\
          & 10.1  & b    & 318  &  \\
          2011 October 29        &  6.7  & v    & 254  & 1.0$\rightarrow$2.6 \\
          &  5.4  & b    & 194  &  \\
          2011 November 2        &  8.5  & v    & 326  & 1.0$\rightarrow$2.2 \\
          &  8.5  & b    & 320  &  \\
          2011 November 13       &  4.4  & v    & 123  & 1.0$\rightarrow$1.9 \\
          &  4.4  & b    & 116  &  \\
          2011 November 14       & 10.4  & v    & 357  & 1.0$\rightarrow$2.7 \\
          & 10.4  & b    & 358  &  \\
          2011 November 15       &  6.4  & v    & 236  & 1.0$\rightarrow$1.8 \\
          &  4.7  & b    & 203  &  \\
          2011 November 26       &  4.8  & v    & 175  & 1.0$\rightarrow$1.5 \\
          &  4.8  & b    & 187  &  \\
          2011 November 27       &  8.6  & v    & 311  & 1.0$\rightarrow$1.7 \\
          &  8.6  & b    & 313  &  \\
          \hline
        \end{tabular}
        \tablefoot{$^a$\;Filter (F), $^b$\; Number of photometric data points
        (NoP)}
\end{table}

\begin{table}
      \centering
      \caption{\label{tab:OBS_CAHA}
      Overview of observation nights for CAHA, OADM, OMNP, and OMCB.}
        \begin{tabular}{l r r r r}
          \hline
          \hline
          Date               & Duration  & F$^a$  & NoP$^b$     & Airmass \\
          & (h) & & & \\ \hline
          \multicolumn{5}{c}{CAHA}  \\
          \hline
          2011 October 22    & 4.1  & $v$, $b$, $y$    & 130  & 1.0$\rightarrow$1.9 \\ 
          2011 October 23    & 3.2  & $v$, $b$, $y$, $d$  & 178  & 1.3$\rightarrow$2.2 \\ 
          2011 October 24    & 3.4  & $v$, $b$, $y$    & 157  & 1.0$\rightarrow$1.9 \\ 
          2011 October 25    & 11.9 & $v$, $b$, $y$    & 462  & 1.0$\rightarrow$2.2 \\ 
          \hline
          \multicolumn{5}{c}{OADM}  \\
          \hline
          2010 September 28      & 3.7   & R    & 279  & 1.0$\rightarrow$1.3 \\
          2011 October 10        & 3.5   & V    & 344  & 1.0$\rightarrow$1.5 \\
          2011 November 23       &   4   & V    & 320  & 1.0$\rightarrow$1.1 \\
          2012 September 24      & 4.2   & V    & 630  & 1.0$\rightarrow$1.0 \\
          2012 October 8         &   4   & V    & 588  & 1.0$\rightarrow$1.1 \\
          \hline
          \multicolumn{5}{c}{OMNP} \\
          \hline
          2011 October 21        & 4.6   & V    & 439  & 1.0$\rightarrow$1.5 \\
          \hline
          \multicolumn{5}{c}{OMCB} \\
          \hline
          2010 August 26         & 6.4   & R    & 162  & 1.0$\rightarrow$2.2 \\
          2010 September 14      & 5.3   & R    & 125  & 1.0$\rightarrow$1.2 \\
          2010 October 20        & 5.6   & R    & 138  & 1.0$\rightarrow$1.7 \\
          2011 October 5         & 7.6   & R    & 195  & 1.0$\rightarrow$1.4 \\
          2011 November 23       & 5.1   & R    & 127  & 1.0$\rightarrow$1.3 \\
          2012 January 1         & 4.5   & V    & 112  & 1.0$\rightarrow$1.2 \\
          2012 January 12        & 5.2   & V    & 130  & 1.0$\rightarrow$1.4 \\
          \hline
        \end{tabular}
        \tablefoot{$^a$\;Filter (F), $^b$\; Number of photometric data points
        (NoP)}
\end{table}


\begin{table}[ht!]
  \centering
  \caption{Technical telescope data: primary mirror diameter, $\O$,
    field of view (FOV), plate scale, and observatory
    location.\label{tab:telescopeData}}
  \begin{tabular}[h]{l l l l l l} 
    \hline
    \hline
    Name & $\O$  & FOV$^a$ & Scale & Location$^b$ \\
    & (m)  &  & (''/pix)  \\ \hline
    OLT     &  $1.2$ & $9'\times 9'$ & 0.158 & G \\
    CAHA    & $2.2$ & $18'\times 18'$ & 0.135 & S  \\
    STELLA  & $1.2$ & $22'\times 22'$ & 0.322 & CI  \\
    OADM    & $0.8$ & $12'\times 12'$ & 0.36 & S \\
    OMNP    & $0.4$ & $21'\times 21'$ & 1.24 & S \\
    OMCB    & $0.3$ & $16'\times 11'$ & 0.62 & S \\ \hline
  \end{tabular}
  \tablefoot{$^a$ Full fields of view are listed.\;$^b$\;Germany
    (G), continental Spain (S), Canary Islands (CI).}
\end{table}

\subsection{Hamburger Sternwarte}
\label{sec:OLTDataReduction}

The Oscar L\"uhning
Telescope\footnote{\url{http:www.hs.uni-hamburg.de/}} (OLT) is located
at the Hamburger Sternwarte, Germany. It is equipped with an Apogee
Alta U9000 charge-coupled device (CCD) camera with guiding system.

Between Nov. 2010 and Oct. 2012, we observed \wasp\ for 29 nights
using the Johnson-Cousins B and R filters (see Table~\ref{tab:OBS_OLT}
for details). The exposure time was between 10 and $40$~s, mainly
depending on the night quality. The airmass ranged from
  principally 1 to 3 only when photometric nights allowed such
  observations. The typical seeing at the Hamburger Sternwarte is
2.5-3~arcsec. Therefore saturation was not an issue. During a
total time of $\sim 150$~h, we obtained $18\,090$ photometric data
points providing both in- and out-of-transit
coverage. Additionally, calibration frames were obtained for each
individual night.

For bias subtraction and flat fielding, we used the \mbox{{\it
    ccdproc}} package in IRAF; aperture photometry was carried out
using IRAF's \mbox{{\it apphot}}. To obtain differential photometry,
we measured unweighted fluxes in \wasp\ and two reference stars using
various aperture radii. The final light curve was produced using the
aperture that minimizes the scatter of the differential light
curve. It is based on the brighter of the two reference stars,
BD+36~488, which is, however, still one magnitude fainter than \wasp.
The remaining reference star was used to obtain control light curves
to ensure that the photometry is based on a proper reference.


\subsubsection{Calar Alto observatory}

The German-Spanish Astronomical Center at Calar Alto (CAHA) is located
close to Almer\'{\i}a, Spain. It is a collaboration between the
Max-Planck-Institut f\"ur Astronomie (MPIA) in Heidelberg, Germany,
and the Instituto de Astrof\'{\i}sica de Andaluc\'{\i}a (CSIC) in
Granada, Spain.

We used the Bonn University Simultaneous CAmera
(BUSCA)\footnote{\url{http://www.caha.es/CAHA/Instruments/BUSCA/intro.html}}
instrument, which is mounted at the 2.2~m telescope. The instrument
allows simultaneous measurements in four different spectral bands. To
reduce read-out time, we exposed only the half-central part of the
CCD. We observed \wasp\ using the Str\"omgren $v$, $b$, and $y$
filters, along with a filter labeled $d$, which is centered at 753 nm
with a FWHM of 30 nm. As BUSCA requires simultaneous read-out of all
CCDs, we used an exposure time of $4$~s, which provides adequate
signal-to-noise ratios in the Str\"omgren bands and avoids saturation
of the source. The photometry obtained by means of the $d$ filter was
discarded due to low signal-to-noise. During our observations, the
seeing ranged between 1 and 1.5~arcsec. In the visible, the extinction
was between 0.15 and 0.2~mag/airmass. Observing for $\sim 23$~h, we
obtained 927 photometric data points per filter. The data reduction
was carried out as described in Sect.~\ref{sec:OLTDataReduction}.


\subsection{Observatorio del Teide}

STELLar Activity (STELLA)\footnote{\url{http://www.aip.de/stella/}}
consists of two fully robotic 1.2~m telescopes, one dedicated to
photometry and the other to spectroscopy \citep{STELLA2010}. The
photometer is a wide-field imager called WiFSIP. It is equipped with a
$4092^2$ 15-$\mu$m pixel back-illuminated CCD.

We observed \wasp\ using STELLA for one month starting at the end of
Oct. 2011. STELLA's optical setup offers a field of view of
$22'\times22'$. However, for the purposes of these observations and
with the main goal of reducing readout times, we used only a
$15'\times15'$ subframe. To obtain quasi-simultaneous multiband
photometry, we alternated between the Str\"omgren $v$ and $b$
filters. In this way, we obtained 2483 photometric measurements with
the $v$ filter and 2234 with the $b$ filter, which equates to $\sim$75
hours per spectral range. Accounting for read-out time, the typical
temporal cadence was of $\sim90$~s. The optics had to be defocused to
avoid saturation of the target.

We carried out the data reduction using ESO-MIDAS. Bias frames were
obtained every night, evening, and morning and combined into a master
bias on a daily basis. Twilight flat-field frames for both the
Str\"omgren $b$ and $v$ filters were obtained approximately every
10~d. Bias subtraction and flat-fielding were performed as usual. We
carried out aperture photometry using SExtractor's MAG AUTO
option. Here, SExtractor computes an elliptical aperture for every
detected object in the field, following its light distribution in x
and y, and scales the aperture width with the SExtractor parameter k,
which we set to 2.6. As a flux calibrator, we used the summed,
unweighted flux of three reference stars, viz., \mbox{BD+36 493},
\mbox{BD+36 487}, and \mbox{BD+36 488}. The background, estimated
locally for each object, was generally low.


\subsection{Primary transit observations}

To increase our sample of primary transit light curves, we used three
telescopes with apertures between $0.3$ and $0.8$~m. To carry out the
observations we defocused the telescopes. In the particular case of
bright sources such as \wasp, long exposures after defocusing can
reduce scintillation noise and flat-fielding errors
\citep[e.g.,][]{Southworth2009,Gillon2009}. In this way, we reached
milli-magnitude precision in all of the primary transit light curves
obtained using small-aperture telescopes.

The Telescopi Joan Or\'o is a fully robotic telescope located at the
Observatori Astron\`omic del
Montsec\footnote{\url{http://www.oadm.cat/en/}} (OADM).  It is
equipped with an FLI Proline~4240 CCD and standard Johnson-Cousins
filters. Observing for five nights distributed over two years, we
collected $19.4$~h of data at a temporal cadence of $\sim 45$~sec.

The Observatori Montcabrer\footnote{\url{http://cometas.sytes.net/}}
(OMCB) is located in Cabrils, Spain. We used its remotely operated
$0.3$~m telescope, which is equipped with an SBIG ST-8 CCD and
standard Johnson-Cousins filters for seven nights distributed over
$1.5$~yr. In total, we collected $\sim 40$~h of data with typical
exposure times of $\sim 120$~s.  Although the observatory is located
in a light-polluted area, a photometric precision of about 1~mmag
could be reached.

The Observatori M\'on Natura
Pirineus\footnote{\url{http://monnaturapirineus.com/en/content/observatory}}
(OMNP) contributed a $0.4$~m telescope equipped with an SBIG
STL-1001E CCD. With this telescope we observed one primary transit
using the Johnson-Cousins V filter.

The data of all the telescopes were corrected for bias and dark current
and were flat-fielded using MaximDL and new calibration images for every
night. The light curves were produced using
Fotodif\footnote{\url{http://www.astrosurf.com/orodeno/fotodif/index.htm}}.
In all cases, the aperture radius was selected such that the
scatter in the out-of-transit sections of each light curve is
minimized. 


\section{\wasp\ as a $\delta$ Scuti star}
\label{Sec3}

The primary transit light curves of \wasp\ are deformed by the host
star's pulsations. This interferes with transit modeling and
therefore with the determination of the orbital and physical parameters of
the system. As the removal of an inappropriate primary transit model
could introduce a spurious signal in the pulsation spectrum of the
star, associated with the planetary orbital period rather than
intrinsic stellar variability, we use only off-transit data points to
determine \waspa's pulsation spectrum.

The pulsation frequency analysis is performed using PERIOD04, a
package intended for the statistical analysis of large astronomical
data sets containing gaps, with single-frequency and
multiple-frequency techniques \citep{LenzBreger2005}. The package
utilizes both Fourier and multiple-least-squares algorithms, which do
not rely on sequential prewhitening or assumptions of white noise.

\subsection{Light curve normalization}

To avoid variations apart from the periodicity that we want to
characterize, the light curves need to be first detrended. Thus, to
study the pulsation spectrum in the high-frequency regime, we
normalize each individual light curve. In this way, we eliminate the
low-frequency signals that might be associated with systematic
effects, such as residual fluctuations due to atmospheric extinction,
that are unrelated to intrinsic stellar variations.

\begin{figure}[ht!]
  \centering
  \includegraphics[width=.5\textwidth]{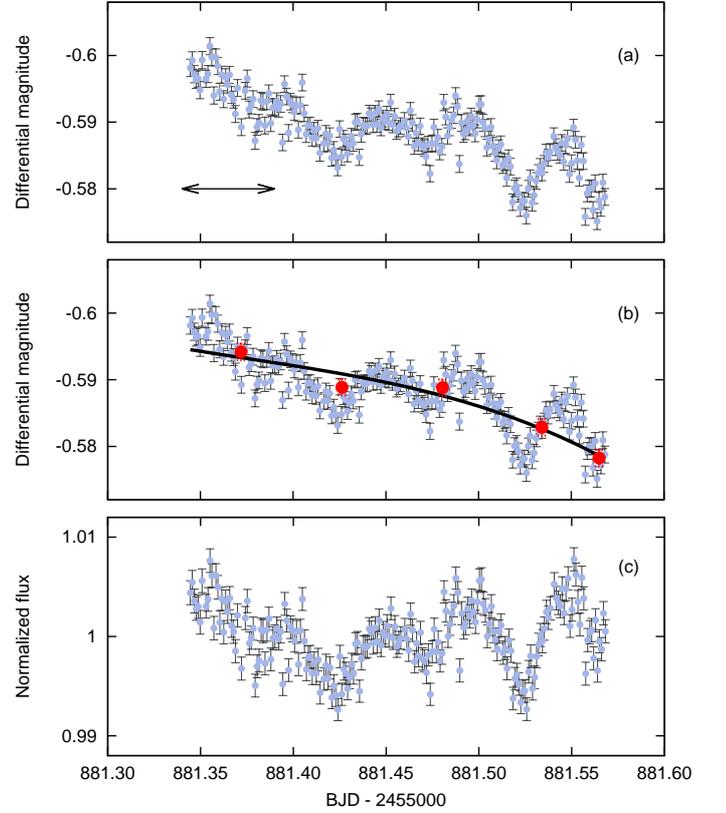}
  \caption{\label{fig:NORMALIZATION} Our normalization procedure: (a)
    Differential light curve in magnitudes, obtained using the STELLA
    telescope and the Str\"omgren $v$ filter. The length of the arrow
    indicates the width of the time-bins. (b) Thick red points: the
    binned time, flux, and photometric error. Continuous black line:
    third-degree polynomial fitted to the binned data points. (c)
    Normalized light curve in flux units.}
\end{figure}

Our procedure is the following: first, we bin the light curves using
time-bins with a duration of $\sim$1.3 hours to ``hide'' the
high-frequency pulsations inside them. Second, we calculate the mean
value of time and flux in the bins and fit a low-order polynomial to
the binned light curves. The degree of the polynomial depends on the
number of available data points and therefore on the duration of the
observing night. Finally, we subtract the fitted polynomial from the
unbinned light curve and convert magnitudes into flux. To ensure a
proper normalization, we visually inspected the results of our
procedure. Figure~\ref{fig:NORMALIZATION} shows a representative
example.

\subsection{Light curve normalization and its relevance for periodogram analysis}

\begin{figure}[ht!]
  \centering
  \includegraphics[width=.5\textwidth]{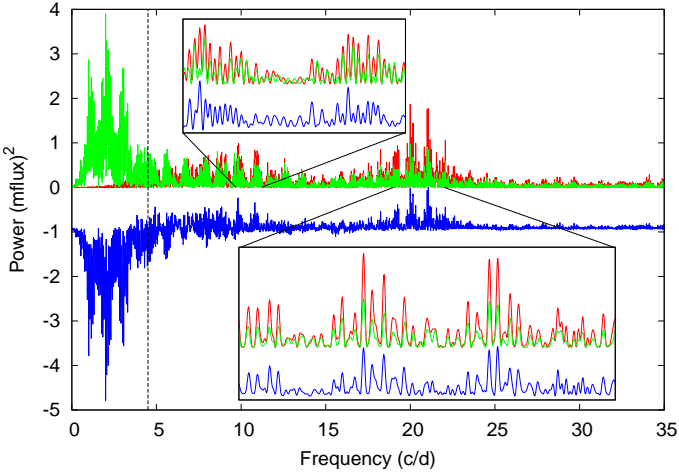}
  \caption{\label{fig:STELLAb_join}Periodogram for STELLA $b$ data for
    the polynomial normalization in red, the alternative normalization
    with a constant in green, and their difference in blue
    (arbitrarily shifted). The vertical dashed line indicates the mean
    duration of the observing nights.}
\end{figure}

To study the impact of the normalization on the high-frequency domain,
we applied an alternative normalization and subtracted only the mean
value of the light curves. We then compared the periodograms obtained
for each data set based on the two normalizations.
Figure~\ref{fig:STELLAb_join} shows the resulting power spectra for
our STELLA data obtained with the Str\"omgren $b$ filter as an
example. The vertical dashed line in the figure indicates the
frequency corresponding to the average length of our observing
nights. The difference between the periodograms shows how the
normalization process affects the power spectrum. The discrepancy is
strongest for frequencies corresponding to periods longer than
the average night length. Beyond that limit, the effect of the
normalization becomes weak. Figure~\ref{fig:STELLAb_join} shows
close-ups of the periodograms around $\nu \sim$ 20 c/d and $\nu \sim$
10 c/d. While the amplitudes of individual peaks change, the structure
of the periodogram and the position of the peaks remain stable.

To verify the stability of the peak positions, we searched both
periodograms for strong peaks and compared their frequency. To
determine the peak positions, we used a Gaussian fit.  Based on
66~peak pairs in the $>10$~c/d regime, we derive a shift of $-0.003
\pm 0.0025$~c/d. At lower frequencies the uncertainty becomes larger,
which is in agreement with the behavior observed in
Figure~\ref{fig:STELLAb_join}. Thus, we conclude that the
normalization does not seriously impede the analysis in the
high-frequency regime.

\subsection{Determination of \wasp's frequency spectrum}
\label{PulsSpec}

Our frequency analysis is based on the STELLA, CAHA, and OLT data. The
latter provide a total temporal coverage of approximately two years,
essentially concentrated, however, in three observing seasons. Each
OLT season was considered separately in our frequency search. We were
left with five data sets obtained in four different spectral filters.
Although the photometric amplitudes of $\delta$~Scuti stars depend on
wavelength, it is possible to combine multifilter data to determine
the frequencies via Fourier methods. Consequently, we combine the data
obtained with the $v$, $B$, and $b$ filters, for which we found the
difference in amplitude values to be statistically insignificant.

We identified the frequencies of the dominating pulsations by
analyzing the combined data set, which provides the cleanest spectral
window and the highest precision in the determined frequencies.

\begin{figure*}[ht!]
  \centering
  \includegraphics[width=.87\textwidth]{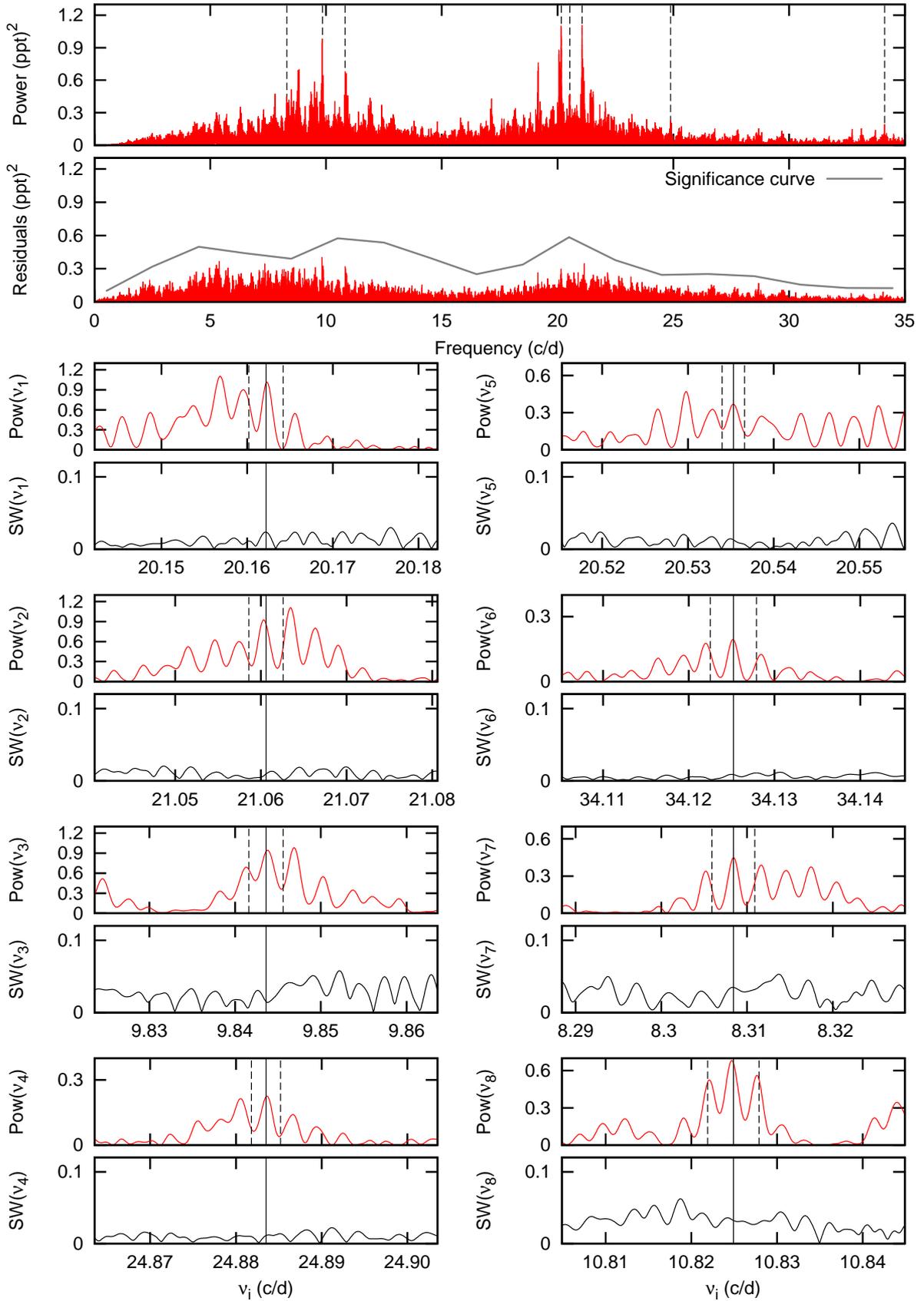}
  \caption{\label{fig:PREWHITENING} Top panels: Periodogram of the
    combined data set and the final residuals in
      part per thousand (ppt). The solid line indicates the
    significance curve. Bottom panels: A closer look at the
    periodogram around the detected frequencies and the spectral
    window (SW). Dashed vertical lines around the peaks indicate our
    error estimates as listed in Table~\ref{tab:FINAL_RES1}.}
\end{figure*}

To estimate the signal-to-noise level of a given pulsation with
amplitude $A_o$, we computed the average amplitude, $\sigma_{res}$,
over a frequency interval with a width of 2~c/d from a periodogram
obtained from the final residuals (see Fig.~\ref{fig:PREWHITENING})
and estimated the amplitude signal-to-noise ratio (ASNR) of each
pulsation as $A_o/\sigma_{res}$. Following \citet{Breger1993}, we
consider a pulsation to be significant when the estimated ASNR of the
periodogram peak is larger than four.
 
The residual power spectrum indicates strong departures from white
noise arising from a potentially highly complex pulsation spectrum
and/or aliasing problems. Nonetheless, all remaining peaks remain
below our significance curve. 

Because we have made annual solutions, the phase and amplitude shifts
from year to year have been taken care of. However, if small and
systematic changes occur from year to year, there will exist smaller
changes within each observing season. Such changes are not taken care of
by PERIOD04 and will lead to close side lobes in the
periodogram. However, since the solution presented in this work does
not contain very close frequencies, these side lobes will not affect
it.

Strong aliasing represents an unavoidable difficulty, leading to 1~c/d
ambiguities and a large number of strong peaks at the 30\% level
(relative to the main peak). This aliasing, combined with a large
number of frequencies, sets the limits of our multifrequency
analysis. To minimize the effect of aliasing and the window function
on our frequency selection, we checked that the significant peaks are
present in all subsamples, where we, however, tolerated lower
amplitude signal-to-noise ratios (ASNR) levels. As the subsamples
comprise only a fraction of the data, the frequencies cannot be
determined with the same accuracy as in the combined sample. We
obtained estimates of the frequency and the uncertainty (see
Sect.~\ref{ErrorTreatment}) from the subsamples and verified that the
results are consistent with the values obtained using the combined
data set. Only if this was the case did we accept a frequency. One
frequency near 7.3~c/d, detected in the combined data set, could not
be found in all individual data sets and is consequently not
included in our analysis.

Altogether, eight frequencies were extracted from the
data. Table~\ref{tab:FINAL_RES1} shows the frequencies, amplitudes,
phases, and associated ASNRs. Additionally, we provide the mean
frequency and error obtained from the subsamples.
Figure~\ref{fig:DATAplusMODEL} shows our pulsation model, plotted over
some of the available off-transit light curves. The displayed
pulsation model is obtained after fitting the phases to each
individual night only. The reasons for this procedure will be given in
Section~\ref{sec:phase-shift}.

\begin{table*}[ht!]
  \centering
  \caption{\label{tab:FINAL_RES1}Parameters of the pulsations with 1-$\sigma$
    errors. Frequency ratios (FR) of the pulsations and the orbit (see Sect.~\ref{SPI}).}
  \begin{tabular}{l l l l l | l  l}
    \hline
    \hline
    \multicolumn{5}{c}{Combined data} & Subsample analysis &  FR \\
     PN   &     $\nu \pm \sigma_{\nu}$ (c/d)  & $A \pm \sigma_A$ (10$^{-3}$) &  $\phi \pm \sigma_{\phi}$ (2$\pi$) & ASNR & $\nu \pm \sigma_{\nu}$ (c/d) &  \\ 
    \hline
     Puls$_1$ &20.16214 $\pm$ 0.00063 & 0.95 $\pm$ 0.03   &  0.5718 $\pm$ 0.0049 & 7.8  & 20.1621 $\pm$ 0.0023  &  24.595 \\
     Puls$_2$ &21.06057 $\pm$ 0.00058 & 0.93 $\pm$ 0.03   &  0.3594 $\pm$ 0.0050 & 7.6  & 21.0606 $\pm$ 0.0023  &  25.691 \\
     Puls$_3$ &~~9.84361 $\pm$ 0.00066 & 0.79 $\pm$ 0.03  &  0.4804 $\pm$ 0.0058 & 6.3  & ~~9.8436 $\pm$ 0.0023 &  12.008 \\
     Puls$_4$ &24.88351 $\pm$ 0.00056 & 0.42 $\pm$ 0.03   &  0.3009 $\pm$ 0.0110 & 4.3  & 24.8835 $\pm$ 0.0017  &  30.355 \\
     Puls$_5$ &20.53534 $\pm$ 0.00057 & 0.71 $\pm$ 0.03   &  0.5207 $\pm$ 0.0065 & 5.9  & 20.5353 $\pm$ 0.0013  &  25.050 \\
     Puls$_6$ &34.12521 $\pm$ 0.00054 & 0.49 $\pm$ 0.03   &  0.5738 $\pm$ 0.0096 & 8.0  & 34.1252 $\pm$ 0.0027  &  41.628 \\
     Puls$_7$ &~~8.30842 $\pm$ 0.00054 & 0.63 $\pm$ 0.03  &  0.5173 $\pm$ 0.0074 & 4.2  & ~~8.3084 $\pm$ 0.0025 &  10.135 \\
     Puls$_8$ &10.82492 $\pm$ 0.00058 & 0.64 $\pm$ 0.03   &  0.9969 $\pm$ 0.0072 & 5.8  & 10.8249 $\pm$ 0.0030  &  13.205 \\
    \hline
  \end{tabular}
\end{table*}

\subsubsection{Analysis of fit quality}

To quantify the improvement in the description provided by our
pulsation model, we calculate the resulting $\chi^2$ values for the
pulsation model and a constant, $\chi^2_{\mathbb{C}}$ and
$\chi^2_{mod}$ and carry out an F-test. In particular, we calculate
the F-statistics using
\begin{equation}
  F = \frac{(\chi_{mod}^2 - \chi_{\mathbb{C}}^2)/(\nu_{mod} -
  \nu_{\mathbb{C}})}{\chi_{\mathbb{C}}^2/\nu_{\mathbb{C}}} \; ,
\end{equation}
\noindent where $\nu$ corresponds to the degrees of freedom,
$\nu_{mod}$ = 14701, and $\nu_{\mathbb{C}}$ = 14725. Formally, we
obtain a $p$-value of 1$\times$10$^{-16}$, indicating that the
pulsation model accounts for a substantial fraction of the light
curve variations.

\begin{figure*}[p]
\centering
  \includegraphics[width=.9\textwidth]{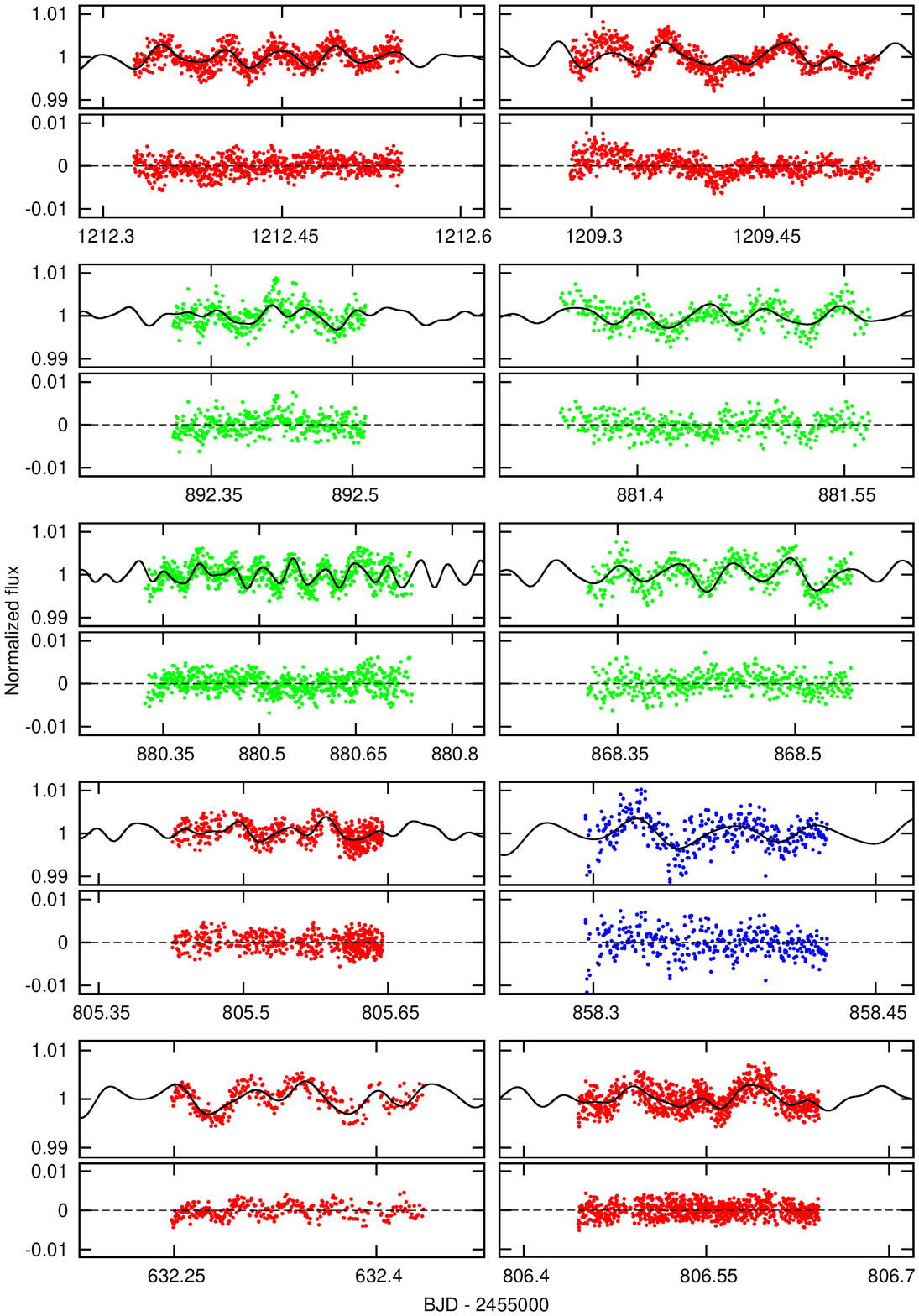}
  \caption{\label{fig:DATAplusMODEL}Exemplary off-transit light curves
    color-coded in red for OLT, green for STELLA, and blue for CAHA,
    overplotted with the pulsation model in black continuous line. Top
    panels: normalized flux. Bottom panels: residuals after the
    pulsation model has been subtracted.}
\end{figure*}

Although our model reproduces the overall stellar pulsation pattern,
the bottom panels of Fig.~\ref{fig:DATAplusMODEL} show flux residuals
that do not behave as random, uncorrelated noise. Such residuals may be
produced by nonsinusoidal pulsations, low-amplitude pulsations not
accounted for in the model or by local changes in the atmospheric
conditions not entirely removed by the differential photometry. At
any rate, the remaining scatter in the data defines the limiting
accuracy achievable in cleaning the primary transits.

\subsection{Error treatment}
\label{ErrorTreatment}

Preliminary error estimates for the frequencies listed in the second
column of Table~\ref{tab:FINAL_RES1} were obtained in two ways. First,
we followed the analytical expressions of \cite{Breger1999a}. Second,
we fitted a Gaussian function to the peaks and used the standard
deviation as our error estimate. To be conservative, we used the larger of
these as frequency-error estimate. As the residuals show correlated
noise, the true uncertainties in our frequencies could be
considerably larger. A deeper discussion of errors is given below.


\subsubsection{Correlated noise and unevenly spaced data for periodogram analysis}

\cite{Montgomery1999} present analytical results of the effect that
random, uncorrelated noise has on a least-squares fit of a sinusoidal,
evenly sampled signal. They provide the following expressions for the
uncertainties:

\begin{equation}
  \sigma_{\nu} = \frac{\sqrt{6}~ \sigma_N}{\pi \sqrt{N} A T}
  \label{eq:UNCORRnoise1} 
\end{equation}
\begin{equation}
  \sigma_{A} = \sqrt{\frac{2}{N}}\sigma_N
  \label{eq:UNCORRnoise2} 
\end{equation}
\begin{equation}
  \sigma_{\phi} = \frac{\sigma_N}{\pi \sqrt{2N}A}\ ,
  \label{eq:UNCORRnoise3} 
\end{equation}

\noindent where $\sigma_{\nu}$, $\sigma_A$, and $\sigma_{\phi}$ are
the standard deviations for a sinusoidal signal with frequency $\nu$,
amplitude $A$, and phase $\phi$. The remaining parameters are $N$ for
the total number of data points, $T$ for the total duration of the
observing campaign, and $\sigma_N$ for the average measurement error
of the data points. If the time series are unevenly sampled and show
correlated noise (e.g., due to atmospheric extinction),
\citet{Montgomery1999} suggest to estimate errors according to

\begin{equation}
  \sigma^2(\omega) = \sigma_o^2(\omega) \cdot A(\omega,D)
  \label{eq:CORRNoise1}
\end{equation}
\begin{equation}
  A(\omega,D) = D \frac{\sqrt{\pi}}{2} e^{-(\frac{D\Delta t \omega}{4})^2} \; ,
  \label{eq:CORRNoise2}
\end{equation}

\noindent where $\sigma_o^2$ is the variance of the parameter for
uncorrelated data sets, given by Eqs.~\ref{eq:UNCORRnoise1},
\ref{eq:UNCORRnoise2}, and \ref{eq:UNCORRnoise3}. Further, $\Delta t$
is the mean exposure time of the data set, $D$ is an estimate of the
number of consecutive correlated data points, and $\omega = 2\pi\nu$
is the angular frequency of the pulsation.

An upper limit for the error can be obtained by maximizing
Eq.~\ref{eq:CORRNoise2}. This occurs when the correlation time is on
the order of the signal period. In this case, we obtain $A = A_{max}
\sim 0.24 P/\Delta t$. Table~\ref{tab:CORRNoise} shows the upper-limit
uncertainty estimates for our pulsation model parameters, obtained by
means of Eq.~\ref{eq:CORRNoise1}. Here, we used the last column of
Table~\ref{tab:FINAL_RES1} as an estimate for the ``uncorrelated''
errors. The estimated upper error limits remain satisfactory to
characterize the pulsations photometrically.

\begin{table}[ht!]
  \centering
  \caption{\label{tab:CORRNoise}Upper limits for the errors of the
    pulsation-associated parameters.}
  \begin{tabular}{l c c c}
    \hline
    \hline
              &   $< \sigma_{\nu} >$  &  $< \sigma_A>$  & $< \sigma_{\phi} >$ \\ 
    \hline
       Puls$_1$ & 0.009 &    0.18 &  0.031 \\
       Puls$_2$ & 0.008 &    0.18 &  0.031 \\  
       Puls$_3$ & 0.021 &    0.38 &  0.078 \\
       Puls$_4$ & 0.007 &    0.15 &  0.058 \\
       Puls$_5$ & 0.008 &    0.18 &  0.041 \\
       Puls$_6$ & 0.004 &    0.11 &  0.036 \\
       Puls$_7$ & 0.020 &    0.49 &  0.116 \\
       Puls$_8$ & 0.016 &    0.35 &  0.087 \\
    \hline
  \end{tabular}
\end{table}


\subsubsection{Photometric errors}

The photometric reduction tasks used in this work neglect systematic
effects and provide statistical measurement errors, which are rather
lower limits to the true uncertainties. We study the impact of the
measurement errors on our frequencies analysis, based on the OLT,
STELLA, and CAHA data (Johnson-Cousins B filter for OLT and
Str\"omgren $v$ for STELLA and CAHA).

In particular, we randomly increase the photometric errors by a factor
of up to two and recalculate the position of the leading peaks and
their respective ASNR. After repeating this procedure 10$^{4}$
times, we analyzed the resulting statistics of peak positions and
ASNR. We find that the observed change in frequency is contained
within the previously derived error. The ASNR decreases but remains
higher than $\sim$4 in all cases. Therefore we conclude that our
frequency analysis is robust against a moderate increase of up to
100\% in the photometric error.


\subsection{Phase-shift analysis}
\label{sec:phase-shift}

Photometry provided by the {\it Kepler} satellite has widely been used
to study the pulsation spectrum and its evolution in $\delta$~Scuti
stars \citep[e.g.,][]{Balona2012a, Southworth2011, Balona2012b,
  Murphy2012}.  Most of the analyzed $\delta$~Scuti stars pulsate in
several modes. For example, \cite{Breger2012} identify 349 frequencies
in the rapidly rotating Sct/Dor star KIC~8054146, for which the
authors even find variations in amplitude and phase.

\begin{figure}
  \centering
  \includegraphics[width=.5\textwidth]{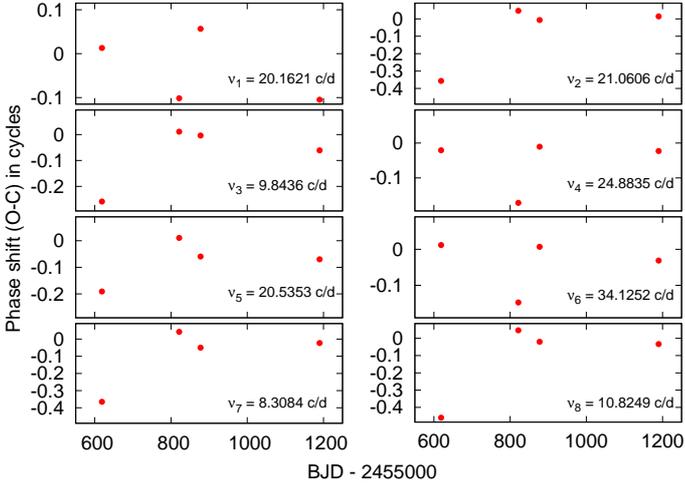}
  \caption{\label{fig:PHASE_SHIFT} Temporal phase evolution of the
    pulsation frequencies.}
\end{figure}

Following the method of \cite{Breger2005LT} to identify amplitude
variations and phase shifts, we divided our off-transit data sets into
four subsets: from BJD$\sim$2455596 to BJD$\sim$2455641 ($\sim$ 1.5
months), from BJD$\sim$2455805 to BJD$\sim$2455837 ($\sim$ 1 month),
from BJD$\sim$2455858 to BJD$\sim$2455896 ($\sim$ 1 month), and from
BJD$\sim$2456162 to BJD$\sim$2456217 ($\sim$ 2 months). This
particular choice avoids including data gaps due to seasonal effects
in the subsamples and therefore limits the impact of aliasing. As
the amplitude of \wasp's pulsations is too low to identify amplitude
variations by means of our photometric data, we focus on phase
shifts. In particular, we fit the phases in each subsample, fixing the
amplitudes and frequencies to the values listed in
Table~\ref{tab:FINAL_RES1}. Figure~\ref{fig:PHASE_SHIFT} shows our
results.  Error bars are on the order of $\sim$0.005 and therefore
rather negligible in the plot.

For all eight detected pulsations, we find a change in phase.  There
are striking similarities between the O-C diagrams of $\nu_2$,
$\nu_3$, $\nu_5$, $\nu_7$, and $\nu_8$, as well as between $\nu_4$ and
$\nu_6$. The largest observed gradient is about $2\times
10^{-3}$~d$^{-1}$, assuming a linear evolution. Clearly, such shifts
must be taken into account in the construction of a pulsation model to
clean the transits.

\subsection{Mode identification}

A particularly interesting question related to the observed
frequencies is their association with specific pulsation modes, i.e.,
the radial order, $n$, degree, $\ell$, and azimuthal number, $m$, of
the underlying spherical harmonic, Y$_{n,\ell}^m$. While the most
reliable method for pulsation mode identification is to analyze the
line-profile variation using high-resolution spectroscopy
\citep{Mathias1997}, our analysis remains limited to photometric
data. Nonetheless, we apply three methods of mode identification based
on photometry.

\subsubsection{Mode identification based on the pulsation constant Q}
\label{sec:Q}

The pulsation constant, $Q$, takes the unique value for any given
pulsation frequency and can be used for mode identification
\citep{Breger1975}.  It is defined by \mbox{$Q =
  P\sqrt{\bar{\rho}/\bar{\rho_{\odot}}}$}, with $P$ being the
pulsation period and $\bar{\rho}$ and $\bar{\rho_{\odot}}$ the mean
densities of the star and the Sun. Two important $Q$-values are
$0.033$~d and $0.026$~d; they correspond to the fundamental and
first radial overtone expected in $\delta$~Scuti stars. Expressing the
densities as a function of the radius and eliminating the radius via
the luminosity, the expression for $Q$ can be recast
\citep{Breger1990}:
\begin{equation}
 \log\left(\frac{Q}{P}\right) = 0.5 \log(g) + 0.1\ M_{bol} + \log(T_{eff}) -
 6.456\ .
\end{equation}

Taking into account the uncertainties in the stellar parameters,
\cite{Breger1990} estimate the uncertainty in the $Q$-value to be
18\%. For \waspa, we adopted \mbox{$\log(g) = 4.3 \pm 0.2$}, \mbox{d =
  116 $\pm$ 16 pc}, and \mbox{$T_{eff}$ = 7430 $\pm$ 100 K}
\citep{CollierCameron2010}. We derived the absolute bolometric
brightness \mbox{$M_{bol} = 2.85 \pm 0.07$} using the expression
\mbox{$M_{bol} = 42.36 - 5\log(R/R_{\odot}) - 10\log(T_{eff})$} \citep{ALLEN}. For
the eight pulsation frequencies in our pulsation model, we derive the
$Q$-values listed in Table~\ref{tab:Qvalues}; errors have been
estimated by error propagation.

\begin{table}[ht!]
  \centering
  \caption{\label{tab:Qvalues} Q values and errors for the eight
    frequencies found in our data.}
  \begin{tabular}{l c}
    \hline
    \hline
    $\nu$ (c/d)                          & Q (d) \\
    \hline
    $\nu_1$ = 20.1621 $\pm$  0.0023 & $Q_1$ = 0.035 $\pm$ 0.008\\
    $\nu_2$ = 21.0606 $\pm$ 0.0023  & $Q_2$ = 0.033 $\pm$ 0.007\\
    $\nu_3$ = ~~9.8436 $\pm$ 0.0023 & $Q_3$ = 0.071 $\pm$ 0.016\\
    $\nu_4$ = 24.8835 $\pm$ 0.0017  & $Q_4$ = 0.028 $\pm$ 0.006\\
    $\nu_5$ = 20.5353 $\pm$ 0.0013  & $Q_5$ = 0.034 $\pm$ 0.007\\
    $\nu_6$ = 34.1252 $\pm$ 0.0027  & $Q_6$ = 0.021 $\pm$ 0.005\\
    $\nu_7$ = ~~8.3084 $\pm$ 0.0025 & $Q_7$ = 0.085 $\pm$ 0.019\\
    $\nu_8$ = 10.8249 $\pm$ 0.0030  & $Q_8$ = 0.065 $\pm$ 0.015\\
    \hline
  \end{tabular}
\end{table}

Comparing the $Q$-values to the ones expected in $\delta$~Scuti stars
\citep[][and references therein]{Breger1998}, we find that $\nu_1$,
$\nu_2$, $\nu_4$, $\nu_5$, and $\nu_6$ are within the range of radial
oscillations. Any further mode identification is not possible via
  $Q$-values.

To illustrate the difficulty of assigning modes accurately only by
means of the pulsation constant, $Q$, we compare our most accurate
$Q_6$-value with the ones theoretically predicted by
\cite{Fitch1981}. The model that best matches the \waspa\ parameters
is labeled ``1.5M21''. Within errors, the following modes correspond
to $Q_6$: first, second, and third harmonic (Table 2A, radial modes);
$p_2$ and $p_3$ (Table 2B, $\ell$ = 1 modes); $p_1$, $p_2$, and $p_3$
(Table 2C, $\ell$ = 2 modes); and $p_1$, $p_2$, and $p_3$ (Table 2D,
$\ell$ = 3 modes). Therefore the only conclusive result is that $Q_6$
corresponds to a p-mode, which is expected for a high-frequency
pulsation.

\subsubsection{The empirical period-luminosity-color relation}

Empirical period-luminosity-color (P-L-C) relations have been studied,
e.g., by \cite{Petersen1998}, \cite{LopezdeCoca1990}, and
\cite{King1991}, among many others.

\citet{Stellingwerf1979} derive a theoretical P-L-C relation
\begin{equation}
  \log P = -0.29 M_{bol} - 3.23 \log(T_{eff}) + \mathbb{C} \; ,
\end{equation}

\noindent where $P$ is the period in days and $\mathbb{C}$ is a
constant equal to 11.96, 11.85, and 11.76 for the fundamental and
first and second harmonics.  Substituting our values for $M_{bol}$ and
$T_{eff}$, yields $\nu_{0,S} = 23.43$~c/d, $\nu_{1,S} = 30.18$~c/d,
and $\nu_{2,S} = 37.13$~c/d, i.e., periods that have not been
identified within our data.

In an observational study, \cite{Gupta1978} finds that a separate
P-L-C relation for each pulsation mode provides a better agreement
with the observations than a general one. The author derived the
following empirical P-L-C relations for the fundamental mode, $F$
(Eq.~\ref{eq1}), and the first, $H1$ (Eq.~\ref{eq2}), and second
harmonic, $H2$ (Eq.~\ref{eq3}):

\begin{align}
   {M_{bol}}_{\pm 0.20} &= -2.83 \log(P_o) - 11.07 \log(T_{eff}) +
  41.61 \label{eq1}\\
   {M_{bol}}_{\pm 0.14} &= -3.57 \log(P_1) - 10.21
  \log(T_{eff}) + 37.13\label{eq2}\\
   {M_{bol}}_{\pm 0.24} &= -2.45 \log(P_2) -
  10.22 \log(T_{eff}) + 38.35\label{eq3} \, .
\end{align}
These relations predicts $\nu_{0,G} = 27.91$~c/d, $\nu_{1,G} =
29.41$~c/d, and $\nu_{2,G} = 45.47$~c/d, again not observed within our
data. At least for \waspa's stellar parameters, the theoretical and
observational relation seem to be mutually inconsistent.

\subsubsection{Multicolor photometry}

In $\delta$~Scuti stars, the photometric amplitude and phase of
pulsations depend on the spectral band. The amplitude and phase of a
given pulsation are determined by the local effective temperature and
cross-section changes, which are defined by the pulsation mode.
Therefore different modes lead to distinguishable modulations in
flux. This allows mode identification to be carried out by means of
multicolor photometry
\citep{Balona1999,DaszynskaDaszkiewicz2003,Dupret2003}.

Frequency Analysis and Mode Identification for AsteroSeismology
(FAMIAS) is a collection of software tools for the analysis of
photometric and spectroscopic time series data \citep{Zima2008}. The
photometry module uses the method of amplitude ratios and phase
differences in different photometric passbands to identify the modes
\citep{Balona1979,Watson1988}. The determination of the l-degrees is
based on static plane-parallel models of stellar atmospheres and on
linear non-adiabatic computations of stellar pulsations. To compute
the theoretical photometric amplitudes and phases, FAMIAS applies the
approach proposed by \cite{DaszynskaDaszkiewicz2002}.

FAMIAS requires the stellar parameters' effective temperature,
$T_{eff}$, surface gravity, $\log g$, and metallicity, [Fe/H], which
we obtained from \citet{CollierCameron2010}.  As additional input to
FAMIAS, we obtained the pulsation frequency, the amplitude, and the
phase for the Str\"omgren $v$ and $b$ bands using our STELLA and CAHA
data; amplitude ratios and phase differences were obtained using
PERIOD04. Figure~\ref{fig:AmpPhase} shows our results for the case of
$\nu_2$, $\nu_4$, and $\nu_5$. The pulsations seem to correspond to
lower-order modes: $\ell$ = 0,1 for $\nu_5$ and $\nu_4$, and $\ell$ =
2,3 for $\nu_2$. A more detailed characterization of these modes
results impossible in our analysis. For the remaining five frequencies
in our model, no reliable information on the associated modes could be
derived.

\begin{figure}[ht!]
  \centering
  \includegraphics[width=.5\textwidth]{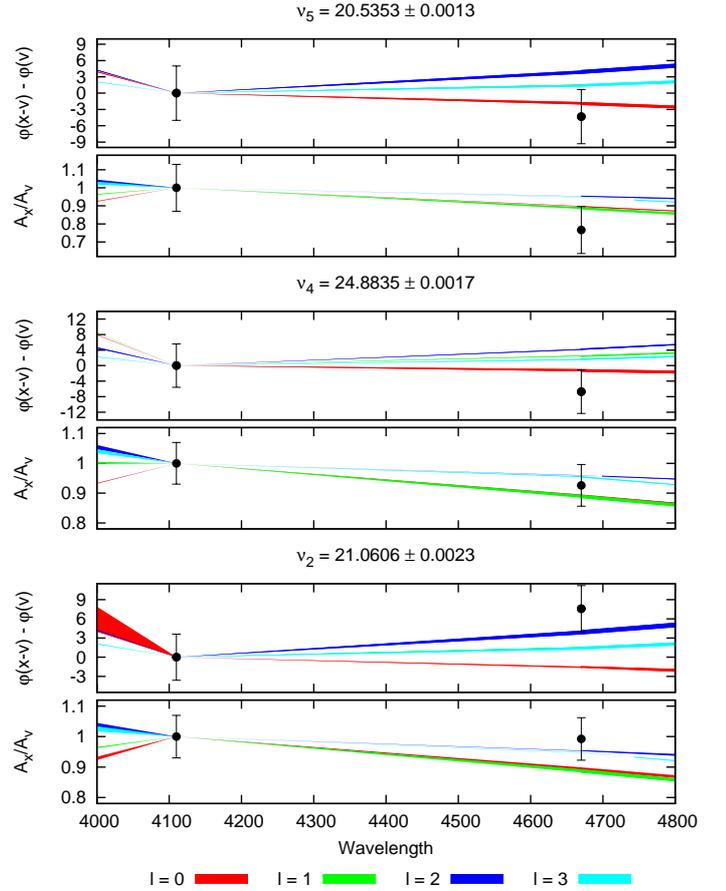}
  \caption{\label{fig:AmpPhase}Amplitude ratios and phase differences
    in degrees relative to the Str\"omgren $v$ filter for \wasp,
    resulting from the three available nights obtained at Calar
    Alto. The filled curves indicate the uncertainty of the
    theoretical prediction due to observational errors in $T_{eff}$
    and $log~g$.}
\end{figure}

\subsubsection{The effect of rotation}


Any asteroseismological study of main-sequence $\delta$ Scuti stars is
not completely fulfilled until stellar rotation is considered
\citep[and references therein]{Goupil2000}. The effects of rotation
over the pulsation spectrum has been theoretically studied
\citep[e.g.,][]{Deupree2011,Deupree2012}, as well as observed
\citep[e.g.,][]{Breger1999b,Breger2005FGVir,Breger2012}. The main
effect of rotation is the splitting of the nonradial mode
frequencies. If such splitting is observed, then the rotation rate of
a star can be determined \citep{CD1991}.

From a purely geometrical argument, stellar rotation affects the
observed frequencies. In an inertial frame, an observer finds that a
frequency is split uniformly according to the azimuthal order $m$:

  \begin{equation}
    \nu_m = \nu_o + \Omega~ m\; ,
  \end{equation} 

\noindent
where $\Omega$ is the angular velocity of the star, $\nu_o$ is the
frequency of the pulsation in the frame rotating with the star, and
$m$ the azimuthal mode. Using this simplified version of mode
splitting \citep[see, e.g.,][for the contribution of Coriolis forces
  to the frequency splitting]{Cowling1949}, we produced the following
analysis: from the eight frequencies that conform to our pulsation
model, we assume that one of them, $\nu_{j,m}$, is the product of mode
splitting. Therefore knowing that the azimuthal order $m$ is
associated with $\nu_{j,m}$ and the rotational period of the star, we
can determine $\nu_o$. With $\nu_o$, we can further calculate the
values of the remaining $\nu_{j,m}$s for a given $\ell$ degree
\mbox{($|m| < \ell$)}, and compare them with our remaining model
frequencies.

Although this approach might sound straight forward, there is no
knowledge of the rotational period of WASP-33A. We therefore assumed
that WASP-33A's $v \sin(i)$ is coincident with the equatorial
velocity. Furthermore, our attempt to identify the nature of the
observed frequencies did not produce any substantial
results. Consequently, for our most accurately identified $\nu_2$
frequency ($\ell$ = 2,3) we assumed all possible $m$ values and found,
through this reasoning, that the remaining observed frequencies were
not the product of mode splitting.

Any further study would require, for instance, a complete mode
identification and the knowledge of the rotational period of the
star. The complexity around mode identification clearly indicates that
Q values, P-L-C relations, rough period ratios, and even poor mode
identification via multicolor photometry cannot be used for mode
identification without further evidence.


\section{Primary transit analysis}
\label{Sec4}

Our data comprise 19 primary transit
observations. Table~\ref{tab:PRIMTRANS} lists, among others, the date
and site of observation, the filter, and a code indicating the transit
coverage of the observation. To determine the orbital parameters, we
focus on the eight primary-transit light curves providing complete
temporal coverage (TC = OIBEO in Table~\ref{tab:PRIMTRANS}).

\cite{Moya2011} report on the detection of an optical companion about
$\sim2$" from \wasp, which could affect our observations through
third-light contamination. Based on the color information ($J_c$,
$H_c$, $K_c$, and FII filters), the authors speculate that it might be
a physical companion of \wasp, for which they estimate an effective
temperature of \mbox{$T_{eff}$ = 3050 $\pm$ 250 K}. As the
third-light contribution provided by such an object is \mbox{$\lesssim
  4\times 10^{-4}$} in all used filters, it can be neglected in our
analysis.

Orbiting a fast rotator in a quasi polar orbit (projected spin-orbit
misalignment \mbox{$\lambda\sim 255^{\circ}$},
\cite{CollierCameron2010}), the transit's light curves may be affected
by gravity darkening, which manifests in a latitudinal dependence of
the stellar effective temperature \citep{vonZeipel1924}. As the
rotational period of \wasp\ is unknown, we estimate it using \mbox{$v
  \sin(i)\sim90$ km s$^{-1}$} and the stellar radius
\mbox{$R_s\sim$1.444 $R_{\odot}$} \citep{CollierCameron2010}. Close to
the system geometry, we estimate the polar-to-equatorial temperature
ratio. Using a gravity-darkening exponent of $\beta$ = 0.18
\citep{Claret1998}, for $g$ the magnitude of the local effective
surface gravity, and $\beta$ the gravity-darkening exponent, following
\cite{Maeder2009}:

\begin{equation}
  T = T_{pole} \frac{g^{\beta}}{g_{pole}^{\beta}} \; ,
\end{equation}

\noindent we estimate that the polar temperature of \wasp\ is $\approx
2.2$\% higher than the equatorial temperature. This is too small to
reproduce the observed transit depth wavelength-dependent
variation. Further, using the primary transit code of
\cite{Barnes2009}, which is adequate for fast rotators, we determine
that the differences in the transit shape observed in the blue and red
bands caused by gravity darkening are on the order of 0.06\% and
therefore negligible in our analysis. Therefore the {\it occultquad}
routine provided by \cite{MAndelAgol2002} is adequate for our transit
modeling.

\begin{table}[ht!]
  \centering
  \caption{\label{tab:PRIMTRANS} Summary of our primary transit
      observations and modeling parameters specifying the filter F,
      the date, the observatory where the transit was acquired, the
      transit coverage (TC), two coefficients, $\beta_{1,2}$, that
      account for correlated noise (Sect.~\ref{photNoise}), and the
      degree of the polynomial used for light curve normalization. A
      description of the transit coding and filters is detailed in the
      footnote of this table.}
  \begin{tabular}{l l l l l l l}
    \hline
    \hline
    F$^{a}$          &       Date        & Obs.  & TC$^{b}$ & $\beta_1$  &  $\beta_2$ & D \\
    \hline
    $B$             & 2011 Sept 24   & OLT    & - IBEO & 2.61 & 2.15 & 0 \\
                    & 2012 Aug 23    & OLT    & OIB - -& 3.14 & 2.47 & 1 \\ 
                    & 2012 Oct 11    & OLT    & OIBE - & 4.55 & 4.15 & 1 \\
                    & 2012 Oct 16    & OLT    & OIBEO  & 4.41 & 3.87 & 1 \\
    \hline
    $V$             & 2011 Oct 10   & OADM   & - IBE -& 3.57 & 2.61 & 0 \\
                    & 2011 Oct 21   & OAPS   & - IBEO & 1.97 & 1.33 & 0 \\
                    & 2012 Jan 1    & RNAV   & OIBEO  & 2.24 & 2.07 & 0 \\
                    & 2012 Jan 12   & RNAV   & OIBEO  & 1.93 & 1.42 & 1 \\
                    & 2012 Sept 24  & OADM   & - - BEO& 4.53 & 3.84 & 2 \\
    \hline
    $R$             & 2010 Aug 26    & RNAV   & OIBEO  & 1.67 & 1.07 & 0 \\  
                    & 2010 Oct 20    & RNAV   & OIBEO  & 1.23 & 1.15 & 1 \\
                    & 2010 Nov 6     & OLT    & OIBEO  & 3.24 & 2.85 & 2 \\
                    & 2011 Oct 5     & RNAV   & OIBEO  & 2.05 & 1.63 & 0 \\
                    & 2011 Nov 23    & RNAV   & OIBEO  & 1.95 & 1.52 & 0 \\
    \hline
    $v$             & 2011 Oct 22    & CAHA   & - IBEO & 1.41 & 1.18 & 0 \\
                    & 2011 Nov 2     & STELLA & OIBE - & 2.36 & 1.95 & 0 \\
    \hline
    $b$             & 2011 Oct 22    & CAHA   & - IBEO & 1.49 & 1.37 & 0 \\
                    & 2011 Nov 2     & STELLA & OIBE - & 2.48 & 2.27 & 0 \\
    \hline
    $y$             & 2011 Oct 22    & CAHA   & - IBEO & 1.52 & 1.06 & 0 \\
    \hline
  \end{tabular}
  \tablefoot{$^a$\; The filter $B$, $V$, and $R$ for Johnson-Cousins and $v$,
  $b$, and $y$ for Str\"omgren filters. $^b$\; The letter code that specifies
  the transit coverage goes as follows: O for ``out of transit before
  ingress'', I for ``ingress'', B for ``bottom'', E for ``egress'', and O for
  ``out of transit, after egress''.}
\end{table}

\subsection{Photometric noise}
\label{photNoise}

Often, the scatter in the light curve is used as a noise estimate. If,
however, correlated noise is present, this method may considerably
underestimate the impact of the scatter on the parameter estimates.
The effect of correlated noise on transit modeling has been studied by
several authors, e.g., \citet{Carter2009, Pont2006}.

While we have identified the significant pulsations in
Sect.~\ref{Sec3}, our analysis has also shown that there is an unknown
number of weak pulsations that we cannot account for in our
modeling. The unaccounted pulsations will manifest in time-correlated
noise in the transit analysis. Therefore a treatment of
time-correlated noise is important in the transit modeling.

To quantify the amplitude of time-correlated noise in our data, we
applied the ``time-averaging method'' proposed by \citet{Pont2006},
which is based on the comparison of the variance of binned and
unbinned residuals. To obtain the residuals, we normalized the transit
light curves by fitting a polynomial to the out-of-transit data and
then subtracted a preliminary transit model. We verified that the
results only slightly depend on the details of the normalization and
transit model.

Subsequently, the residual light curves were divided into M bins of
equal duration. Each bin contains N data points. As our data are not
always equally spaced, we applied a mean value for the number of data
points per bin. In the absence of red noise, the expectation value of
the variance of the unbinned residuals, $\sigma_1$, is related to the
variance of the binned residuals, $\sigma_N$, according to
\citep[cf.][Eq.~36]{Carter2009}
\begin{equation}
  \sigma_N = \sigma_1 \sqrt{\frac{M}{N\,(M-1)}} \; .
\end{equation}
This may now be compared with a variance estimate, $\sigma'_N$,
derived from the binned residuals
\begin{equation}
  \sigma'_N = \sqrt{\frac{1}{M}\sum_{i = 1}^{M}(\overline{\mu} - \mu_i)^2}
  \;\;\;\mbox{with}\;\;\;\overline{\mu} = \frac{1}{M} \sum\limits_{j=1}^{M}
  \mu_j\; .
\end{equation}
If correlated noise is present, then $\sigma'_N$ will differ from
$\sigma_N$ by a factor $\beta_N$, which estimates the strength of
correlated noise. A proper estimator, $\beta$, may be found by
averaging $\beta_N$ over a range $\Delta n$ corresponding to the most
relevant timescale. To account for the correlated noise in a
conventional white-noise analysis, the individual photometric errors
are enlarged by a factor of $\beta$. If there is no prior information,
this leaves the parameter estimates unaffected and enlarges the errors
by a factor $\beta$.

In the case of our transit analysis, the relevant timescale is the
duration of ingress or egress, which is $\sim 16$~minutes for
\wasp. In Table~\ref{tab:PRIMTRANS} we show the resulting $\beta$
factors. In a first step, we deliberately ignored our results derived
in Sect.~\ref{Sec3} and treated the light curves as if we had no
knowledge on the pulsations. In the thus derived $\beta_1$ values, all
pulsations show up as correlated noise. In a second step, we
subtracted the pulsation model derived in Sect.~\ref{Sec3} and
determined the $\beta_2$ values. Taking into account the pulsation
model always yields a better, that is, smaller $\beta$ factor,
indicating that the model accounts for a substantial fraction of the
time correlation.

\subsection{Polynomial order of transit light curve normalization}

In transit modeling, normalization of the light curves is crucial. To
normalize the transit light curves, we fit polynomials with an order
between zero and four to the out-of-transit data and determine the
order, $k$, that minimizes the Bayesian information criterion,
\mbox{BIC = $\chi^2$ + k ln N}.

The order of the resulting optimal polynomial is listed in
Table~\ref{tab:PRIMTRANS}. According to our modeling, a constant or
linear model is sufficient to normalize the transit in all but two
cases, where a quadratic normalization is required. Finally, we
visually inspected the resulting light curves to ensure a proper
normalization.

\subsection{Transit modeling}

\citet{CollierCameron2010} detect the planetary ``shadow'' of
\waspb\ in the line profile of \waspa\ during transit. Their
spectroscopic time series analysis reveals that the planet traverses
the stellar disk at an inclination angle incompatible with
90$^{\circ}$. As the inclination is less affected by parameter
correlations in the spectroscopic analysis, we impose a Gaussian prior
on the inclination. In particular, we use the value of $i = 87.67 \pm
1.8$~deg obtained by \citet{CollierCameron2010} in their combined
photometric and spectroscopic analysis.

In our analysis, we fixed the linear and quadratic limb-darkening
coefficients, $u_1$ and $u_2$, to the values listed in
Table~\ref{tab:LDCs}. We obtained these values, which consider the
stellar parameters \mbox{$\log(g) = 4.5$}, \mbox{$T_{eff} = 7500$},
and \mbox{$[Fe/H] = 0.1$}, from \citet{Claret2011}. Therefore we are
left with the following free parameters: the mid-transit time, $T_o$,
the orbital period, $Per$, the semimajor axis in stellar radii,
$a/R_s$, the orbital inclination, $i$, and the planet-to-star radius
ratio, $p = R_p/R_s$. As previously mentioned, we used complete
primary transits {\it only} to determine the best-fit orbital
parameters.

\begin{table}[ht!]
  \centering
  \caption{\label{tab:LDCs} Linear ($u_1$) and quadratic ($u_2$)
    limb-darkening coefficients.}
  \begin{tabular}{c c c}
    \hline
    \hline
    Filter & $u_1$ & $u_2$ \\
    \hline
    J-C B           &  0.3561 &  0.3625 \\
    J-C V           &  0.2725 &  0.3535 \\
    J-C R           &  0.1954 &  0.3511 \\
    Str v            &  0.3828 &  0.3678 \\
    Str b            &  0.3612 &  0.3512 \\
    Str y            &  0.2795 &  0.3533 \\
    \hline
  \end{tabular}
\end{table}

To obtain the parameter estimates, their errors, and mutual
dependence, we sample from the posterior probability distribution
using an Markov chain Monte Carlo (MCMC) approach. For the parameters
$a$, $p$, $Per$, and $T_o$, we defined uniform priors covering a
reasonable range. Errors are given as 68.3\% highest probability
density (HPD) credibility intervals. To carry out MCMC sampling, we
used Python routines of
\texttt{PyAstronomy}\footnote{\url{http://www.hs.uni-hamburg.de/DE/Ins/Per/Czesl
    a/}\\ \url{PyA/PyA/index.html}}, which provide a convenient
interface to fit and sample algorithms implemented in the PyMC
\citep{Patil2010} and SciPy \citep{Jones2001} packages.

In a first attempt to fit the transits, we ignore the pulsations and
fit only the transit light curves. In our approach, the complete
transit light curves are fitted simultaneously using the model of
\citet{MAndelAgol2002}. We note that we also fitted the coefficients of
the normalizing polynomial, whose degree remains, however, fixed to
that listed in Table~\ref{tab:PRIMTRANS}. Our best-fit solutions,
which are obtained after $5\times10^5$ iterations, are given in
Table~\ref{tab:param_NYpuls}.

In a second attempt, we combine the primary transit model with the
pulsation model with frequencies and amplitudes fixed to the values
listed in Table~\ref{tab:FINAL_RES1}. In Sect.~\ref{sec:phase-shift},
we demonstrate that there is a temporal evolution in the
phases. Therefore the phases have been considered free parameters in
our modeling. However, we did not allow them to take arbitrary values
but restricted the allowed range to the limiting cases derived in
Sect.~\ref{sec:phase-shift}. For instance, the phase of the first
frequency, $\nu_1$, could not deviate by more than $0.1$~cycle from
the mean value (cf. Fig.~\ref{fig:PHASE_SHIFT}).

The results are presented in the lower part of
Table~\ref{tab:param_NYpuls}; in Fig.~\ref{PrimTransModels} we show
the 19 primary transit light curves and the best-fit model.
Interestingly, the parameters derived using this more elaborate
approach are consistent with those obtained ignoring the
pulsations. Taking into account the pulsation model does, however,
improve the uncertainties in the parameter estimates with respect to
the regular primary transit-fitting approach.

The values derived in our analysis are broadly consistent with those
derived previously by \citet{CollierCameron2010} and
\citet{Kovacs2013}. While we find a slightly smaller semimajor axis
than \citet{CollierCameron2010}'s, the planet-to-star radius ratio and
the inclination are compatible. \citet{Kovacs2013} find an $8-10$\%
larger radius ratio and a slightly lower inclination. A homogeneous
study of all the primary transits available in the bibliography
escapes the purpose of this work. However, we believe that the small
differences in the orbital parameters observed by different authors
might be the product of an inadequate normalization of the primary
transits or an insufficiently correlated noise treatment.

\begin{table}[ht!]
  \centering
  \caption{\label{tab:param_NYpuls} Parameters obtained by
    \citet{CollierCameron2010} and \citet{Kovacs2013}, our best-fit
    result obtained from primary transit modeling: : first, ignoring
    pulsations, and second, taking into account pulsations.}
  \begin{tabular}{l l}
    \hline
    \hline
    Parameter         &  Value        \\
    \hline 
    \multicolumn{2}{c}{\citet{CollierCameron2010}}  \\ \hline            
    $a$ ($R_s$)       &  3.79 $\pm$ 0.02   \\ 
    $i$ ($^{\circ}$)  &  87.67 $\pm$ 1.81   \\
    $p$ ($R_p/R_s$)   &  0.1066 $\pm$ 0.0009   \\
    $Per$ (days)      &  1.2198669 $\pm 1.2\times10^{-6}$  \\
    \hline
    \multicolumn{2}{c}{\citet{Kovacs2013}}  \\ \hline            
    $a$ ($R_s$)       &  3.69 $\pm$ 0.01   \\ 
    $i$ ($^{\circ}$)  &  86.2 $\pm$ 0.2   \\
    $p$ ($R_p/R_s$)   &  0.1143 $\pm$ 0.0002   \\
    \hline
    \multicolumn{2}{c}{Transit fit ignoring pulsations}  \\ \hline
    $a$ ($R_s$)       &  3.69$\pm$ 0.04                    \\ 
    $i$ ($^{\circ}$)  &  88.17 $\pm$ 1.53                 \\
    $p$ ($R_p/R_s$)   &  0.1052 $\pm$ 0.0008             \\
    $Per$ (days)      &  1.2198667 $\pm 1.5\times10^{-6}$   \\
    T$_o$ (\bjdtdb)   &  $2455507.5225 \pm 0.0004$         \\
    \hline
    \multicolumn{2}{c}{Transit fit accounting for pulsations}  \\ \hline
    $a$ ($R_s$)       & 3.68 $\pm$ 0.03                    \\      
    $i$ ($^{\circ}$)   & 87.90 $\pm$ 0.93                  \\     
    $p$ ($R_p/R_s$)   & 0.1046 $\pm$ 0.0006          \\      
    $Per$ (days)     & 1.2198675 $\pm$ 1.1$\times$10$^{-6}$  \\
    $T_o$ (\bjdtdb)   & 2455507.5222 $\pm$ 0.0003     \\   \hline
  \end{tabular}
\end{table}

\begin{figure*}[p]\ContinuedFloat*
  \centering
  \includegraphics[width=.9\textwidth]{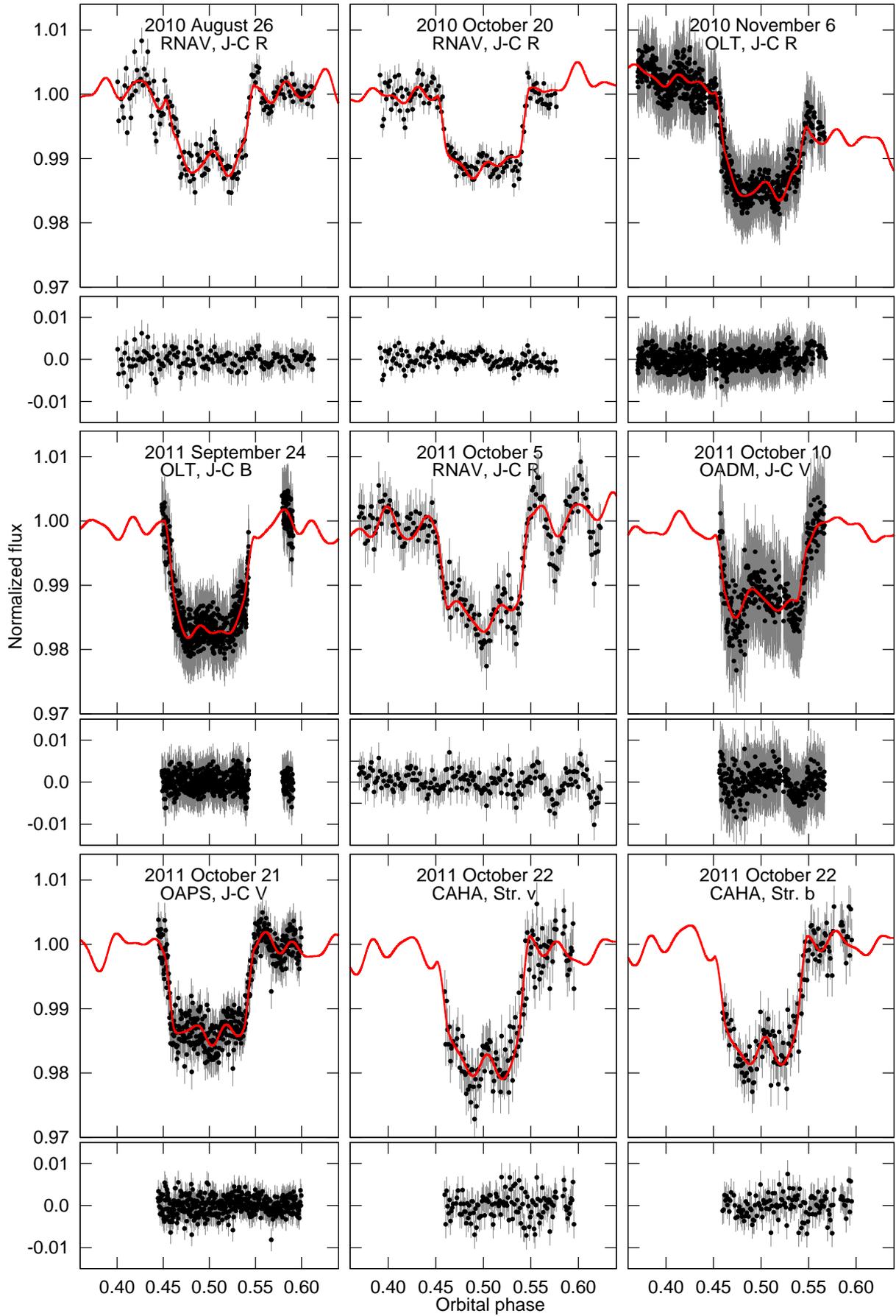}
  \caption{\label{PrimTransModels} Top panels: The 19 primary transits
    in black points, along with the photometric error bars accounting
    for correlated noise, cf. Sec.~\ref{photNoise}. Overplotted in
    continuous red line is the best-fitted primary transit model
    modulated by the host star pulsations and the low-order
    normalization polynomial. Bottom panels: residuals.}
\end{figure*}

\begin{figure*}[p]\ContinuedFloat
  \centering
  \includegraphics[width=.9\textwidth]{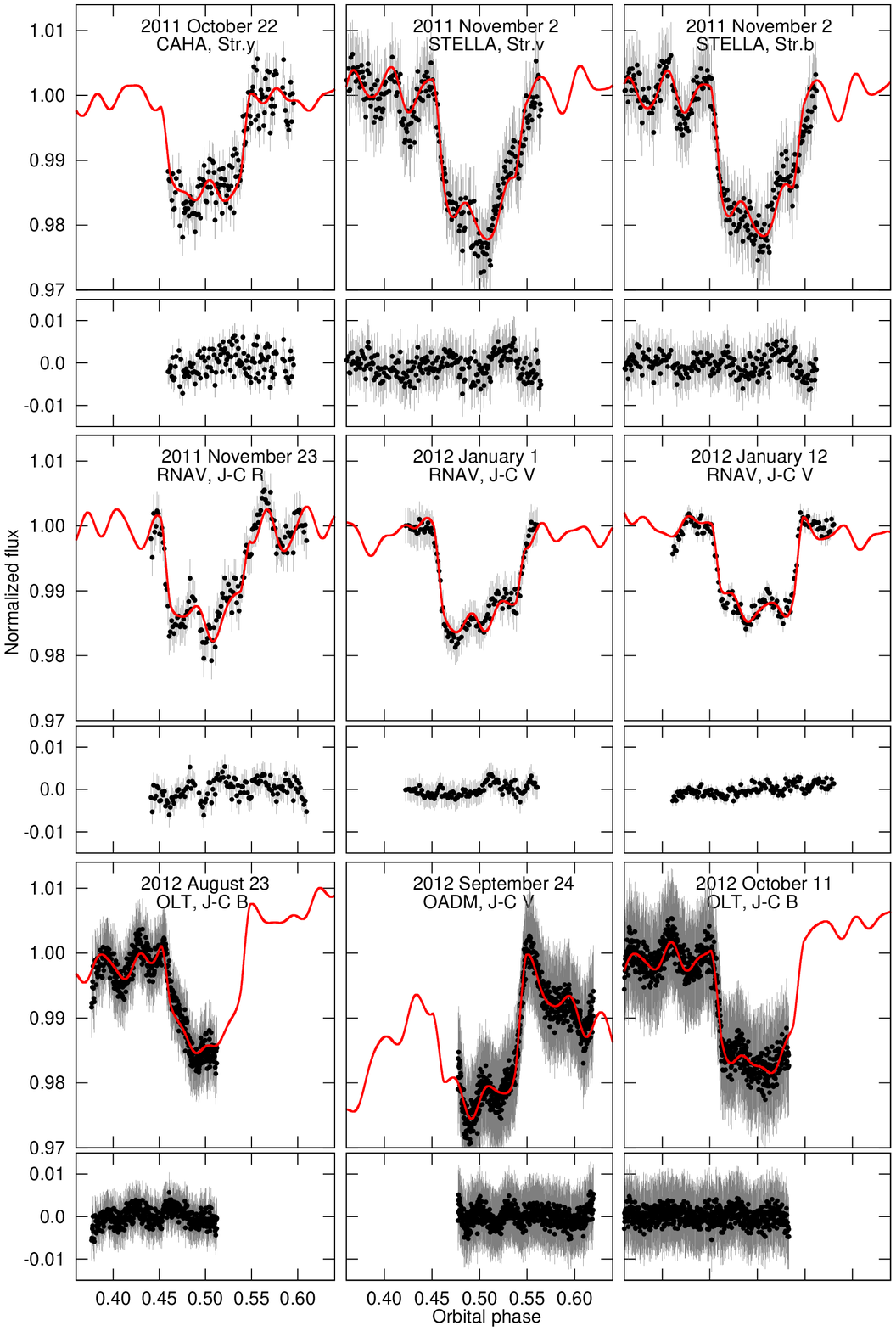}
  \caption{See Figure~\ref{PrimTransModels}}
\end{figure*}

\begin{figure*}[ht!]\ContinuedFloat
  \centering
  \includegraphics[width=.4\textwidth]{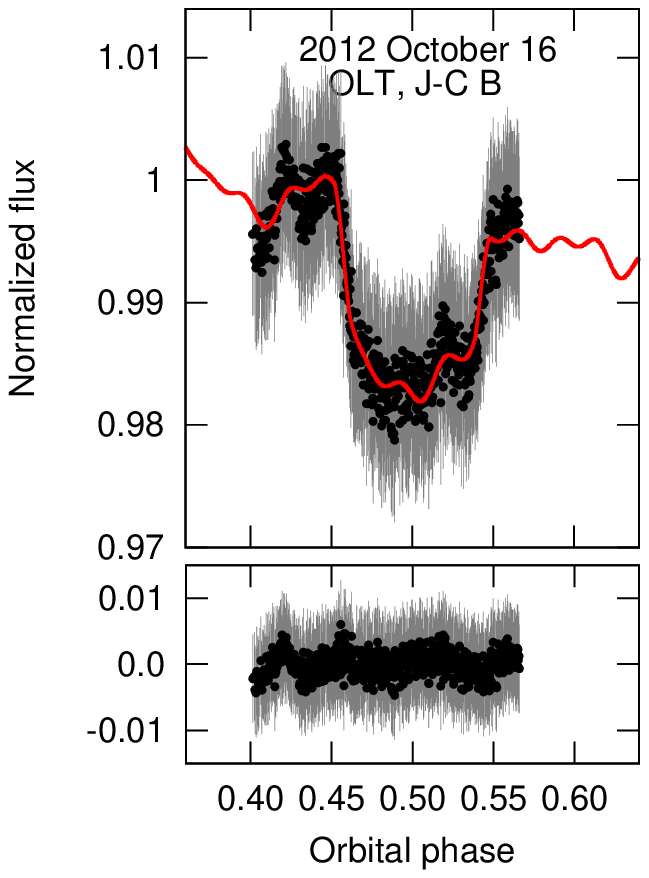}
  \caption{See Figure~\ref{PrimTransModels}}
\end{figure*}

\subsubsection{Impact of the pulsations on the transit fits}

To better understand the effect of the pulsations on the transit fits,
we fit the 19 primary transits individually and study the
behavior of $a$, $i$, and $p$. We carry out the fit one more time
first ignoring the pulsations and then taking them into account via
our pulsation model. During the fit, the ephemeris were fixed to the
corresponding values in Table~\ref{tab:param_NYpuls}. The outcomes are
based on $5\times10^5$ iterations of the MCMC sampler; they are given
in Table~\ref{tab:indParam}.

\begin{table*}[ht!]
  \centering
  \caption{\label{tab:indParam} Results of individual transit
    fits. Complete primary transits are indicated with an *.}
  \begin{tabular}{l l l l l | l}
    \hline
    \hline
    Date           & F    &      a ($R_s$)          &      i ($^{\circ}$)          
    & p & $\hat{p}$  \\
    \hline
    \multicolumn{6}{c}{Pulsations ignored} \\ \hline
 2010 Aug. 26 *    & R & 3.88 $\pm$ 0.12  & 87.74 $\pm$ 1.77 & 0.0999 $\pm$ 0.0026 & 0.0988 $\pm$ 0.0027 \\ %
 2010 Oct. 20 *   & R & 3.72 $\pm$ 0.06  & 88.05 $\pm$ 1.60 & 0.1006 $\pm$ 0.0016 & 0.1004 $\pm$ 0.0016 \\ %
 2010 Nov. 6 *   & R & 3.69 $\pm$ 0.08  & 87.62 $\pm$ 1.78 & 0.1095 $\pm$ 0.0019 & 0.1071 $\pm$ 0.0023 \\ %
 2011 Sept. 24  & B & 3.63 $\pm$ 0.06  & 88.07 $\pm$ 1.62 & 0.1239 $\pm$ 0.0021 & 0.1231 $\pm$ 0.0019 \\
 2011 Oct. 5 *    & R & 3.50 $\pm$ 0.10  & 87.73 $\pm$ 1.76 & 0.1145 $\pm$ 0.0033 & 0.1131 $\pm$ 0.0032 \\ %
 2011 Oct. 10    & V & 3.44 $\pm$ 0.17  & 87.76 $\pm$ 1.73 & 0.1052 $\pm$ 0.0035 & 0.1041 $\pm$ 0.0031 \\
 2011 Oct. 21    & V & 3.48 $\pm$ 0.06  & 87.94 $\pm$ 1.69 & 0.1111 $\pm$ 0.0018 & 0.1101 $\pm$ 0.0017 \\
 2011 Oct. 22    & v & 3.68 $\pm$ 0.10  & 87.72 $\pm$ 1.79 & 0.1241 $\pm$ 0.0028 & 0.1238 $\pm$ 0.0027 \\
 2011 Oct. 22    & b & 3.56 $\pm$ 0.10  & 87.61 $\pm$ 1.82 & 0.1223 $\pm$ 0.0026 & 0.1217 $\pm$ 0.0025 \\
 2011 Oct. 22    & y & 3.61 $\pm$ 0.11  & 87.74 $\pm$ 1.75 & 0.1137 $\pm$ 0.0033 & 0.1134 $\pm$ 0.0031 \\
 2011 Nov. 2    & v & 3.92 $\pm$ 0.17  & 87.36 $\pm$ 1.87 & 0.1311 $\pm$ 0.0032 & 0.1287 $\pm$ 0.0032 \\
 2011 Nov. 2    & b & 3.92 $\pm$ 0.21  & 87.34 $\pm$ 1.92 & 0.1271 $\pm$ 0.0035 & 0.1248 $\pm$ 0.0036 \\
 2011 Nov. 23 *  & R & 3.56 $\pm$ 0.09  & 87.64 $\pm$ 1.80 & 0.1101 $\pm$ 0.0030 & 0.1088 $\pm$ 0.0030 \\ %
 2012 Jan. 1$^{st}$ *&V & 3.43 $\pm$ 0.09  & 87.73 $\pm$ 1.77 & 0.1113 $\pm$ 0.0033 & 0.1093 $\pm$ 0.0035 \\ %
 2012 Jan. 12 *   & V & 3.70 $\pm$ 0.07  & 87.93 $\pm$ 1.69 & 0.1058 $\pm$ 0.0018 & 0.1061 $\pm$ 0.0017 \\ %
 2012 Aug. 23     & B & 4.21 $\pm$ 0.18  & 86.94 $\pm$ 2.07 & 0.1174 $\pm$ 0.0024 & 0.1093 $\pm$ 0.0026 \\
 2012 Sep. 24    & V & 3.83 $\pm$ 0.14  & 87.77 $\pm$ 1.75 & 0.1167 $\pm$ 0.0037 & 0.1176 $\pm$ 0.0032 \\
 2012 Oct. 11    & B & 3.48 $\pm$ 0.12  & 87.69 $\pm$ 1.79 & 0.1115 $\pm$ 0.0032 & 0.1131 $\pm$ 0.0032 \\
 2012 Oct. 16 *   & B & 3.63 $\pm$ 0.10  & 87.71 $\pm$ 1.76 & 0.1098 $\pm$ 0.0031 & 0.1097 $\pm$ 0.0031 \\ %
    \hline
     \multicolumn{6}{c}{Pulsations taken into account} \\ \hline
 2010 Aug. 26 *    & R & 3.74  $\pm$ 0.07  & 87.80  $\pm$ 1.71  & 0.1011  $\pm$ 0.0015 & 0.1008  $\pm$ 0.0015 \\ %
 2010 Oct. 20 *   & R & 3.73  $\pm$ 0.06  & 88.11  $\pm$ 1.59  & 0.1046  $\pm$ 0.0014 & 0.1046  $\pm$ 0.0014 \\ %
 2010 Nov. 6 *  & R & 3.61  $\pm$ 0.07  & 87.83  $\pm$ 1.69  & 0.1050  $\pm$ 0.0016 & 0.1050  $\pm$ 0.0016 \\ %
 2011 Sept. 24  & B & 3.64  $\pm$ 0.06  & 88.20  $\pm$ 1.57  & 0.1194  $\pm$ 0.0024 & 0.1186  $\pm$ 0.0020 \\
 2011 Oct. 5 *    & R & 3.49  $\pm$ 0.09  & 87.62  $\pm$ 1.73  & 0.1143  $\pm$ 0.0024 & 0.1112  $\pm$ 0.0024 \\ %
 2011 Oct. 10    & V & 3.59 $\pm$ 0.11   & 87.88  $\pm$ 1.70  & 0.1046 $\pm$ 0.0039 &  0.1030 $\pm$  0.0028 \\
 2011 Oct. 21    & V & 3.54  $\pm$ 0.06  & 87.79  $\pm$ 1.75  & 0.1103  $\pm$ 0.0013 & 0.1098  $\pm$ 0.0012 \\
 2011 Oct. 22    & v & 3.49  $\pm$ 0.13  & 87.57  $\pm$ 1.84  & 0.1236  $\pm$ 0.0008 & 0.1229  $\pm$ 0.0025 \\
 2011 Oct. 22    & b & 3.33  $\pm$ 0.10  & 87.56  $\pm$ 1.83  & 0.1235  $\pm$ 0.0024 & 0.1216  $\pm$ 0.0024 \\
 2011 Oct. 22    & y & 3.52  $\pm$ 0.13  & 87.65  $\pm$ 1.77  & 0.1143  $\pm$ 0.0023 & 0.1136  $\pm$ 0.0022 \\
 2011 Nov. 2    & v & 3.79  $\pm$ 0.17  & 87.03  $\pm$ 1.00  & 0.1212  $\pm$ 0.0016 & 0.1296  $\pm$ 0.0025 \\
 2011 Nov. 2    & b & 3.90  $\pm$ 0.25  & 87.14  $\pm$ 1.96  & 0.1277  $\pm$ 0.0034 & 0.1249  $\pm$ 0.0032 \\
 2011 Nov. 23 *  & R & 3.53  $\pm$ 0.08  & 87.64  $\pm$ 1.83  & 0.1126  $\pm$ 0.0023 & 0.1093  $\pm$ 0.0022 \\ %
 2012 Jan. 1$^{st}$ * &V & 3.54  $\pm$ 0.09  & 87.76  $\pm$ 1.75  & 0.1082  $\pm$ 0.0026 & 0.1077  $\pm$ 0.0026 \\%
 2012 Jan. 12 *   & V & 3.73  $\pm$ 0.07  & 87.80  $\pm$ 1.69  & 0.1034  $\pm$ 0.0012 & 0.1031  $\pm$ 0.0011 \\ %
 2012 Aug. 23     & B & 4.00  $\pm$ 0.12  & 87.29  $\pm$ 1.87  & 0.1157  $\pm$ 0.0019 & 0.1135  $\pm$ 0.0019 \\
 2012 Sep. 24     & V & 3.89  $\pm$ 0.09  & 87.70  $\pm$ 1.75  & 0.1206  $\pm$ 0.0033 & 0.1186 $\pm$ 0.0030 \\
 2012 Oct. 11    & B & 3.27  $\pm$ 0.09  & 87.42  $\pm$ 1.85  & 0.1245  $\pm$ 0.0023 & 0.1256  $\pm$ 0.0023 \\
 2012 Oct. 16 *   & B & 3.64  $\pm$ 0.10  & 87.79  $\pm$ 1.71  & 0.1054  $\pm$ 0.0027 & 0.1075  $\pm$ 0.0030 \\ %

    \hline
  \end{tabular}
\end{table*}

To study the impact of the pulsation model on the individual
parameters, we scrutinized the ratio of the derived values. In
particular, we focused on the eight complete transits whose parameters
can be determined most reliably. For the ratios of values determined
with pulsations considered in the model (wp) and neglected pulsations
(pn), we obtained $a_{wp}/a_{pn} = 1.0 \pm 0.02$, $i_{wp}/i_{pn} =
1.000 \pm 0.001$, and $p_{wp}/p_{pn} = 0.99\pm 0.03$. These numbers
indicate that, on average, the parameter estimates remain unaffected
by taking into account the stellar pulsations. Regarding individual
fits, the expected deviation amounts to $0.08$~R$_s$ in the semimajor
axis, $0.1^{\circ}$ in the inclination, and $3\times 10^{-3}$ in
$p$. Clearly, the relative uncertainty is largest in the semimajor
axis and the radius ratio, $p$. Also taking into account the transits
with incomplete observational coverage, we obtain numbers that are
comparable but with larger uncertainties. It is worth mentioning that
the ``outlying'' orbital parameters, which are presented in
Table~\ref{tab:indParam}, are the product of incomplete primary
transit fitting. Therefore we believe that primary transit
normalization might play a fundamental role in the determination of
such parameters.

\subsection{Wavelength dependence of the planet-to-star radius ratio}

Our data comprise transit observations from the blue to the red
filter. To check whether a dependence of the planet-to-star radius
ratio on the wavelength can be identified, we fixed all parameters but
the radius ratio to the values listed in Table~\ref{tab:param_NYpuls}
and fitted only the radius ratio, $\hat{p}$, for each individual
transit. The resulting $\hat{p}$ values, which are based on the
pulsation-corrected light curves, are listed in the last column of
Table~\ref{tab:param_NYpuls}. We verified that we obtain comparable
results if the pulsations are not considered.

\begin{figure}[ht!]
  \centering
  \includegraphics[width=.5\textwidth]{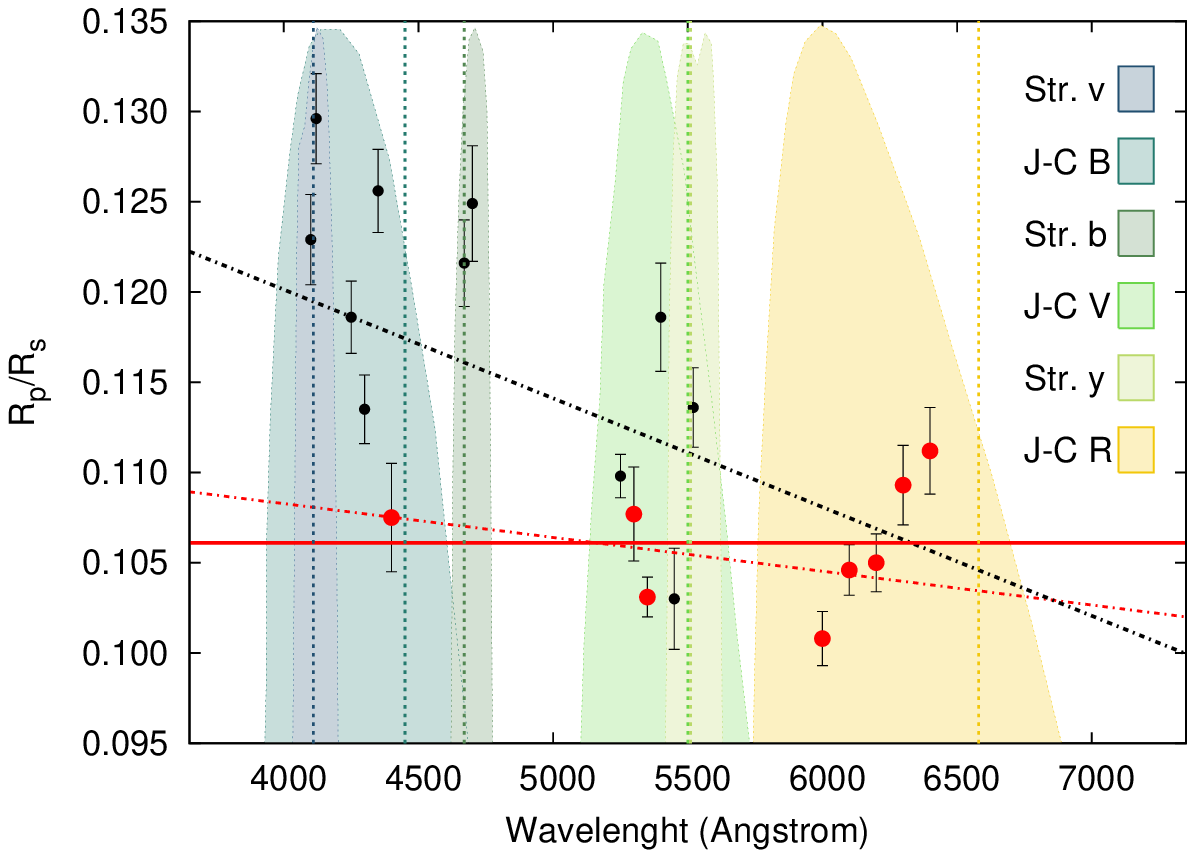}
  \caption{\label{fig:RpRs_vs_lambda} Planet-to-star ratio $\hat{p} =
    R_p/R_s$ obtained from complete light curves (red points) and
    incomplete ones (black points) when the pulsations have been
    accounted for in the model fitting. Vertical color-dashed lines
    indicate the central wavelength of each filter. The dashed/dotted
    black line shows the best-fitting linear model to the 19 $Rp/Rs$
    data points. In red and considering complete primary transits
    only, the dashed/dotted line shows the low-significant
    wavelength-dependent trend, while the continuous line accounts for
    the mean radius ratio.}
\end{figure}

Figure~\ref{fig:RpRs_vs_lambda} shows $\hat{p}$ as a function of
wavelength. The red points mark the transits with full observational
coverage, while the black points were derived from transits with
incomplete temporal coverage (see Table~\ref{tab:PRIMTRANS}).  In the
fit, we used the central wavelength of the filters, which are
indicated by vertical, dashed lines in the plot. To produce a less
crowded figure, the data points are artificially shifted from the
central wavelength.

The values for $\hat{p}$, which are derived from the complete
transits, show a wavelength-dependent trend, but only with marginal
significance. These values are also consistent with a constant radius
ratio of $R_P/R_S = $0.1061 $\pm$ 0.0031, in concordance with the mean
value and the standard deviation of the eight data points. Both the
linear and constant models produce comparable $\chi^2_{red}$ values of
1.9 and 1.8, respectively. Hence, it is unclear which model best fits
the data.  If all transit observations are taken into account, the
data indicate a decrease of 0.65\%/1000\AA\ in the planet-to-star
radius ratio with wavelength. Formally, the correlation coefficient
between radius-ratio and wavelengths amounts to $r \sim 0.7$. We
caution, however, that the observed trend may be feigned by an
inappropriate transit normalization because many transits lack full
observational coverage.

\citet{Kovacs2013} also notice a substantial difference in the transit
depth derived from an observation in H$\alpha$ and a simultaneously
obtained $J$-band light curve (their Fig.~9). In particular, the
$J$-band transit is shallower, which would be consistent with the
wavelength trend. However, the signature observed through the
H$\alpha$ line may significantly differ from those observed in
broadband filters because the H$\alpha$ line is affected by strong
chromospheric contributions.

\section{Discussion}

\subsection{The stellar pulsation spectrum}

The pulsation spectrum of \wasp\ has been studied by several authors,
who report a wealth of frequencies (see Table~\ref{tab:All_Puls}). All
studies find pulsations with amplitudes on the order of 1 mmag, which
is compatible with our results.

\cite{Deming2012} observed \wasp\ during two secondary transits in the
$K_s$ band using the 2.1~m telescope at Kitt Peak National Observatory
and for another two nights using the Spitzer telescope. All
observations were performed during secondary transits. Their frequency
analysis was carried out for individual nights. Their first three
frequencies ($21.1$, $20.2$, and $9.8$~c/d) are compatible with our
$\nu_1$, $\nu_2$, and $\nu_3$ (see Table~\ref{tab:FINAL_RES1}).  Also,
\cite{DeMooij2013} observed secondary transits of \waspb\ in the
$K_s$ band for two nights, each lasting $\sim 5$~h. Although the
frequencies they report are within the range of values we find, the
values are numerically inconsistent with our results.

\cite{Herrero2011} observed for nine nights mainly during primary
transit using Johnson-Cousins $R$ filter. They report pulsation
frequencies, which might correspond to our $\nu_1$.  \cite{Smith2011}
carried out observations during one secondary transit using an S[III]
narrow-band filter centered at $9077$~\AA. The observations were
performed for almost nine hours during a single night lacking photometric
conditions. Among the pulsation frequencies they find,
$21.6\pm0.6$~c/d and $34.3\pm0.4$~c/d likely correspond to our $\nu_2$
and $\nu_6$.

The most extensive pulsation analysis is carried out by
\cite{Kovacs2013}. It is based on four photometric datasets, including
that of \cite{Herrero2011}. \citet{Kovacs2013} report two frequencies
that are compatible with our $\nu_1$ and $\nu_2$. The $\sim 15.2$~c/s
frequency does not show up in our analysis.
 
Among the different data sets, pulsations with frequencies around
$\sim$21~c/d and $\sim$20~c/d that correspond to our most pronounced
frequencies $\nu_1$ and $\nu_2$ consistently occur. Other pulsation
frequencies are only found in some cases.  We note, however, that the
data were acquired in different spectral bands, mostly in the
infrared, where pulsations are expected to be lower in amplitude. The
residuals after subtracting our pulsation model clearly indicate the
presence of further low-amplitude pulsations, which might correspond
to those found in previous studies. Additionally, our phase shift
analysis in Sect.~\ref{sec:phase-shift} has shown that the pulsation
spectrum might be intrinsically variable.  The amplitudes of the
pulsations found by the various studies are all on the order of
$1$~mmag, which is compatible with our results. As the amplitudes are
intrinsically small and furthermore wavelength dependent, we refrain
from a detailed comparison of the derived numbers.

\begin{table}[ht!]
  \centering
  \caption{ From top to bottom, the evolution of the reported
      frequencies and amplitudes for WASP-33's pulsation spectrum.
  \label{tab:All_Puls}}
  \begin{tabular}{l l}
    \hline
    \hline
    Frequency & Amplitude \\
    (c/d) & (mmag) \\
    \hline
 \multicolumn{2}{c}{\citet{Herrero2011}} \\ \hline
 21.004 $\pm$ 0.004 & 0.98 $\pm$ 0.05 \\
 21.311 $\pm$ 0.004 & $\sim$0.86 \\
 \hline  
 \multicolumn{2}{c}{\citet{Smith2011}} \\ \hline
 26.9 $\pm$ 0.4 & 1.479 $\pm$ 0.069 \\
 18.8 $\pm$ 0.6 & 0.567 $\pm$ 0.134 \\
 34.3 $\pm$ 0.4 & 0.766 $\pm$ 0.115 \\
 21.6 $\pm$ 0.6 & 0.605 $\pm$ 0.105 \\ \hline
    \multicolumn{2}{c}{\citet{Deming2012}} \\
    \hline
 $\sim$21.1 & $\sim$1.3 \\  
 $\sim$20.2 & $\sim$2.3   \\
 $\sim$9.8 & $\sim$1.6 \\
 $\sim$26.6 & $\sim$2.1  \\
 $\sim$11.4 & $\sim$2.1  \\
 \hline  
 \multicolumn{2}{c}{\citet{DeMooij2013}$^a$} \\ \hline
 22.5 $\pm$ 0.1 & 0.95 $\pm$ 0.04$^{N1}$ \\
 33.3$^{N1}$ & 0.41 $\pm$ 0.04 \\ 
 27.3 $\pm$ 0.2 & 0.56 $\pm$ 0.06$^{N2}$ \\
 33.2$^{N2}$ & 0.17 $\pm$ 0.05 \\
 22.0$^{N1}$ & 0.11 $\pm$ 0.06 \\
 17.1$^{N1}$ & 0.13 $\pm$ 0.06 \\
 \hline  
 \multicolumn{2}{c}{\citet{Kovacs2013}$^b$} \\ \hline
 15.21643 $\pm$ 0.00004 & 0.758 $\pm$ 0.085$^{HN}$ \\
 20.16229 $\pm$ 0.00004 & 0.733 $\pm$ 0.080$^{HN}$ \\
 21.06339 $\pm$ 0.00004 & 0.719 $\pm$ 0.078$^{HN}$ \\
 15.21517 $\pm$ 0.00001 & 0.477 $\pm$ 0.054$^{H+F}$ \\
 20.16230 $\pm$ 0.00001 & 0.739 $\pm$ 0.053$^{H+F}$ \\
 21.06346 $\pm$ 0.00001 & 0.728 $\pm$ 0.049$^{H+F}$ \\
 \hline  
 \multicolumn{2}{c}{ This work} \\ \hline
 20.1621 $\pm$ 0.0023  & 1.03 $\pm$ 0.03 \\
 21.0606 $\pm$ 0.0023  & 1.01 $\pm$ 0.03 \\
~~9.8436 $\pm$ 0.0023  & 0.86 $\pm$ 0.03 \\
 24.8835 $\pm$ 0.0017  & 0.45 $\pm$ 0.03 \\
 20.5353 $\pm$ 0.0013  & 0.77 $\pm$ 0.03 \\
 34.1252 $\pm$ 0.0027  & 0.53 $\pm$ 0.03 \\
~~8.3084 $\pm$ 0.0025  & 0.68 $\pm$ 0.03 \\
 10.8249 $\pm$ 0.0030  & 0.69 $\pm$ 0.03 \\ 
\hline
  \end{tabular}

\tablefoot{$^a$\; Following the nomenclature of \citet{DeMooij2013}, the
superscripts ${N1}$ and ${N2}$ refer to ``Night I'' and ``Night II''.
$^b$\; The superscript ${HN}$ stands for ``HATNet''
and ${H+F}$ for ``HATNet+FUP''.
}
\end{table}

In our analysis, we identify eight significant pulsation
frequencies. Although the pulsation spectrum is probably much more
complex than that, the amplitudes of the pulsations are intrinsically
low and more data is required to characterize the components in
further detail. We show that the pulsation phases vary in time with a
gradient, $dp$, of up to $|dp| \lesssim 2\times 10^{-3}$~d$^{-1}$,
assuming a linear evolution. This suggests that the amplitudes and
frequencies also show temporal variability. However, our data do not
allow to verify this.

We find that most of the detected frequencies are likely associated
with low-order p-modes. We attempt to further identify the modes using
Q values, empirical P-L-C relations, and amplitudes and phases of
multicolor photometry. However, we find the detected frequencies to
be largely incompatible with all these relations. We argue that this
is not uncommon \citep[e.g.][]{Breger2005}.


\subsection{Transit modeling}

In their analysis of photometric follow-up data, \citet{Kovacs2013}
find a persistent ``hump'' in the residuals obtained after subtracting
the transit model shortly after mid-transit time. While our residuals
clearly show unaccounted pulsations, we do not see any such hump
recurring at the same phase. Therefore we find no evidence for a
persistent structure on the stellar surface like, e.g., a spot belt as
suggested by \citet{Kovacs2013}.

Although our transit modeling is consistent with a constant
star-to-planet radius ratio with respect to wavelength, there may be a
slight trend indicating a decrease in the radius ratio as the
wavelength increases. Although this would be compatible with the
results of \citet{Kovacs2013}, who find that the radius ratio in the
$J$ band is smaller than that observed in H$\alpha$, we caution that
the formation of the H$\alpha$ line may be different from that
observed in broadband filters -- a caveat already mentioned by
\citet{Kovacs2013}. Based on the currently available data, we conclude
that a constant planet-to-star radius ratio seems most likely.

\subsection{Star-planet interaction}
\label{SPI}

\citet{CollierCameron2010} report on a nonradial pulsation at a
frequency of $\sim 4$~c/d, which might be tidally induced by the
planet. Unfortunately, this frequency lies outside the sensitivity
range of our analysis.

In close binary systems, tidal interaction affects stellar
oscillations \citep[e.g.,][]{Cowling1941, Savonije2002,
  Willems2003}. In particular, \cite{Hambleton2013} studied a
short-period binary system that presents $\delta$ Scuti pulsations and
tidally excited modes. In addition to the already known
commensurability between the pulsation frequencies and the orbital
period of the system, the authors found that the spacing between the
detected p-modes was an integer multiple of the system's orbital
frequency. Although it is clear that the nature of WASP-33 does not
resemble a short-period eccentric binary system, in order to analyze
star-planet interaction we search for commensurability of the detected
pulsation frequencies with the planetary orbital period and
investigated the spacings between them.

As the exact rotation period of the host star \waspa\ remains unknown,
it is not entirely clear how exactly the planet affects the stellar
surface in the frame of the star; in particular, the period at which
the planet affects the same surface element is unknown for the larger
fraction of the stellar surface. However, as the planet orbits in a
highly tilted, nearly polar orbit, the stellar poles experience a
periodic force with a period identical to once and twice the planetary
orbital period. When the planet crosses a pole, the effective gravity
on both poles is lowered due to the planetary gravity and orbital
motion; the effect is, however, not the same on both poles.

Using our best-fit orbital period of $1.2198675$~d, we express the
pulsation frequencies in terms of the orbital frequency of the
planet. The result is presented in the last column of
  Table~\ref{tab:FINAL_RES1}, where the ratios of the pulsations and
  the orbit are displayed. We expect the error in the pulsation
frequencies to be considerably larger than those in the orbital
period. The closest commensurability is found for the $9.8436$~c/d,
which corresponds to $12.008$ times the orbital frequency.

To assess the significance of such a result, we carry out a
Monte-Carlo simulation. In particular, we randomly generate eight
frequencies between $8$ and $34$, i.e., in the approximate range of our
detected pulsations. The frequencies are dropped out from a
  uniform distribution. We then calculate the associated ratio based
on the orbital frequency and finally search for the best
match. After $50\,000$ runs, we find that the cumulative probability
distribution for the minimum distance from an integer frequency ratio
is given by

\begin{equation*}
  F(d_{min}) = 1 - e^{-\frac{d_{min}}{0.060934}} \; ,
\end{equation*}

\noindent obtained after fitting our Monte-Carlo results with an
  exponential decay. Using this relation, we find that the
probability of detecting at least one of the ratios as close or closer
than $0.008$~c/d to an integer ratio to be $12$\%. The ratio may
indeed be an integer, considering the error in the frequency
determination (see Sect.~\ref{Sec3}). Nevertheless, we note that our
phase-shift analysis also revealed variability in the frequency
spectrum, which we find hard to reconcile with a simple picture of
tidally excited pulsations.

Additionally, we found that the spacing between the frequencies cannot
be described by harmonics of the orbital period of the system. In
fact, the best case scenario is given by $\nu_5$ and
$\nu_7$. Considering our best-fit orbital period, the departure from
an integer number is ten times their estimated error.

Therefore we conclude that there is no evidence for a direct relation
between any of our pulsation frequencies and the planetary orbital
period. 



\section{Conclusions}
In this work, we obtained and analyzed an extensive set of photometric
data of the hottest known star hosting a hot Jupiter, \wasp. The data
cover both in- and out-of-transit phases and are used to study the
pulsation spectrum and the primary transits.

In particular, our out-of-transit data provide $\sim$3~times more
temporal coverage than the \cite{Kovacs2013} data set, which is the
most extensive among those listed in Table~\ref{tab:All_Puls}. In
addition, our data set is the only one that comprises dedicated
out-of-transit photometric coverage to study the stellar pulsations in
detail, along with multicolor and simultaneous observations to study
the nature of the modes.

A comprehensive study of the pulsation spectrum of \wasp\ reveals, for the
first time, eight significant frequencies. Additionally, some of the
frequencies found seem to be consistent with previous reports. Along
with the associated amplitudes and phases, we construct a pulsation
model that we use to correct the primary transit light curves with
the main goal of redetermining the orbital parameters by means of
pulsation-clean data.

In our transit modeling, we find the system parameters broadly
consistent with those reported by \citet{CollierCameron2010} and
\citet{Kovacs2013}. Interestingly, the derived parameter values are
hardly affected by taking into account the pulsations in the modeling
although the errors decrease. This statement clearly depends on the
total number of observed transits and the stability of the
p-modes. Thus further observations of primary transit events of
WASP-33 will be required to support or overrule this remark.

One possible explanation of the behavior of the orbital parameters
with respect to the pulsations of the host could be that the
associated amplitudes, at least in the high-frequency range that our
studies focus on, are small in nature. Furthermore, our extensive
primary transit observations, which we obtained in different filter
bands, allow us to notice a decrease in the planet-to-star radius
ratio with wavelength. This decrease has also been observed by other
authors. Simultaneous multiband photometry of primary transits of
\wasp\ will help to better constrain this dependency.

Considering that our work was produced using fully ground-based
observations, we were able to provide an extensive study of the
pulsation spectrum of this unique $\delta$ Scuti host star. This, in
turn, has helped to better comprehend how much pulsations affect the
determination of system parameters.

\begin{acknowledgements}

C. von Essen acknowledges funding by the DFG in the framework of RTG
1351, Dr. Andres Moya and Prof. Rafael Garrido for discussing issues
on $\delta$ Scuti stars, and the anonymous referee for her/his
improvements on the draft. E. H. and I. R. acknowledge financial
support from the Spanish Ministry of Economy and Competitiveness
(MINECO) and the ``Fondo Europeo de Desarrollo Regional'' (FEDER)
through grant AYA2012-39612-C03-01. The Joan Or\'o Telescope (TJO) of
the Montsec Astronomical Observatory (OAdM) is owned by the
Generalitat de Catalunya and operated by the Institute for Space
Studies of Catalonia (IEEC). We further thank Ramon Naves for
operating the 0.3 m telescope at Observatori Montcabrer, Thomas
Granzer for his support on STELLA observations and data reduction, and
the anonymous referee for his/her report.

\end{acknowledgements}

\bibliographystyle{aa}
\bibliography{vonEssen_W33}

\begin{thebibliography}{70}
\expandafter\ifx\csname natexlab\endcsname\relax\def\natexlab#1{#1}\fi

\bibitem[{{Allen}(1973)}]{ALLEN}
{Allen}, C.~W. 1973, {Astrophysical quantities}

\bibitem[{{Baglin} {et~al.}(1973){Baglin}, {Breger}, {Chevalier}, {Hauck}, {Le
  Contel}, {Sareyan}, \& {Valtier}}]{Baglin1973}
{Baglin}, A., {Breger}, M., {Chevalier}, C., {et~al.} 1973, \aap, 23, 221

\bibitem[{{Balona} {et~al.}(2012{\natexlab{a}}){Balona}, {Breger}, {Catanzaro},
  {Cunha}, {Handler}, {Ko{\l}aczkowski}, {Kurtz}, {Murphy}, {Niemczura},
  {Papar{\'o}}, {Smalley}, {Szab{\'o}}, {Uytterhoeven}, {Christiansen},
  {Uddin}, \& {Stumpe}}]{Balona2012b}
{Balona}, L.~A., {Breger}, M., {Catanzaro}, G., {et~al.} 2012{\natexlab{a}},
  \mnras, 424, 1187

\bibitem[{{Balona} \& {Evers}(1999)}]{Balona1999}
{Balona}, L.~A. \& {Evers}, E.~A. 1999, Delta Scuti Star Newsletter, 13, 26

\bibitem[{{Balona} {et~al.}(2012{\natexlab{b}}){Balona}, {Lenz}, {Antoci},
  {Bernabei}, {Catanzaro}, {Daszy{\'n}ska-Daszkiewicz}, {di Criscienzo},
  {Grigahc{\`e}ne}, {Handler}, {Kurtz}, {Marconi}, {Molenda-{\.Z}akowicz},
  {Moya}, {Nemec}, {Pigulski}, {Pricopi}, {Ripepi}, {Smalley}, {Su{\'a}rez},
  {Suran}, {Hall}, {Kinemuchi}, \& {Klaus}}]{Balona2012a}
{Balona}, L.~A., {Lenz}, P., {Antoci}, V., {et~al.} 2012{\natexlab{b}}, \mnras,
  419, 3028

\bibitem[{{Balona} \& {Stobie}(1979)}]{Balona1979}
{Balona}, L.~A. \& {Stobie}, R.~S. 1979, \mnras, 189, 649

\bibitem[{{Barnes}(2009)}]{Barnes2009}
{Barnes}, J.~W. 2009, \apj, 705, 683

\bibitem[{{Breger}(1979)}]{Breger1979}
{Breger}, M. 1979, \pasp, 91, 5

\bibitem[{{Breger}(1990)}]{Breger1990}
{Breger}, M. 1990, Delta Scuti Star Newsletter, 2, 13

\bibitem[{{Breger}(1998)}]{Breger1998}
{Breger}, M. 1998, in Astronomical Society of the Pacific Conference Series,
  Vol. 135, A Half Century of Stellar Pulsation Interpretation, ed. P.~A.
  {Bradley} \& J.~A. {Guzik}, 460

\bibitem[{{Breger}(2005)}]{Breger2005LT}
{Breger}, M. 2005, in Astronomical Society of the Pacific Conference Series,
  Vol. 335, The Light-Time Effect in Astrophysics: Causes and cures of the O-C
  diagram, ed. C.~{Sterken}, 85

\bibitem[{{Breger} \& {Bregman}(1975)}]{Breger1975}
{Breger}, M. \& {Bregman}, J.~N. 1975, \apj, 200, 343

\bibitem[{{Breger} {et~al.}(2012){Breger}, {Fossati}, {Balona}, {Kurtz},
  {Robertson}, {Bohlender}, {Lenz}, {M{\"u}ller}, {L{\"u}ftinger}, {Clarke},
  {Hall}, \& {Ibrahim}}]{Breger2012}
{Breger}, M., {Fossati}, L., {Balona}, L., {et~al.} 2012, \apj, 759, 62

\bibitem[{{Breger} {et~al.}(1999{\natexlab{a}}){Breger}, {Handler}, {Garrido},
  {Audard}, {Zima}, {Papar{\'o}}, {Beichbuchner}, {Li}, {Jiang}, {Liu}, {Zhou},
  {Pikall}, {Stankov}, {Guzik}, {Sperl}, {Krzesinski}, {Ogloza}, {Pajdosz},
  {Zola}, {Thomassen}, {Solheim}, {Serkowitsch}, {Reegen}, {Rumpf},
  {Schmalwieser}, \& {Montgomery}}]{Breger1999a}
{Breger}, M., {Handler}, G., {Garrido}, R., {et~al.} 1999{\natexlab{a}}, \aap,
  349, 225

\bibitem[{{Breger} {et~al.}(2005{\natexlab{a}}){Breger}, {Lenz}, {Antoci},
  {Guggenberger}, {Shobbrook}, {Handler}, {Ngwato}, {Rodler}, {Rodriguez},
  {L{\'o}pez de Coca}, {Rolland}, \& {Costa}}]{Breger2005FGVir}
{Breger}, M., {Lenz}, P., {Antoci}, V., {et~al.} 2005{\natexlab{a}}, \aap, 435,
  955

\bibitem[{{Breger} {et~al.}(2005{\natexlab{b}}){Breger}, {Lenz}, {Antoci},
  {Guggenberger}, {Shobbrook}, {Handler}, {Ngwato}, {Rodler}, {Rodriguez},
  {L{\'o}pez de Coca}, {Rolland}, \& {Costa}}]{Breger2005}
{Breger}, M., {Lenz}, P., {Antoci}, V., {et~al.} 2005{\natexlab{b}}, \aap, 435,
  955

\bibitem[{{Breger} {et~al.}(1999{\natexlab{b}}){Breger}, {Pamyatnykh},
  {Pikall}, \& {Garrido}}]{Breger1999b}
{Breger}, M., {Pamyatnykh}, A.~A., {Pikall}, H., \& {Garrido}, R.
  1999{\natexlab{b}}, \aap, 341, 151

\bibitem[{{Breger} {et~al.}(1993){Breger}, {Stich}, {Garrido}, {Martin},
  {Jiang}, {Li}, {Hube}, {Ostermann}, {Paparo}, \& {Scheck}}]{Breger1993}
{Breger}, M., {Stich}, J., {Garrido}, R., {et~al.} 1993, \aap, 271, 482

\bibitem[{{Breger} \& {Stockenhuber}(1983)}]{Breger1983}
{Breger}, M. \& {Stockenhuber}, H. 1983, Hvar Observatory Bulletin, 7, 283

\bibitem[{{Campbell} \& {Wright}(1900)}]{Campbell1900}
{Campbell}, W.~W. \& {Wright}, W.~H. 1900, \apj, 12, 254

\bibitem[{{Carter} \& {Winn}(2009)}]{Carter2009}
{Carter}, J.~A. \& {Winn}, J.~N. 2009, \apj, 704, 51

\bibitem[{{Christensen-Dalsgaard} \& {Berthomieu}(1991)}]{CD1991}
{Christensen-Dalsgaard}, J. \& {Berthomieu}, G. 1991, {Theory of solar
  oscillations}, ed. A.~N. {Cox}, W.~C. {Livingston}, \& M.~S. {Matthews},
  401--478

\bibitem[{{Christian} {et~al.}(2006){Christian}, {Pollacco}, {Skillen},
  {Street}, {Keenan}, {Clarkson}, {Collier Cameron}, {Kane}, {Lister}, {West},
  {Enoch}, {Evans}, {Fitzsimmons}, {Haswell}, {Hellier}, {Hodgkin}, {Horne},
  {Irwin}, {Norton}, {Osborne}, {Ryans}, {Wheatley}, \&
  {Wilson}}]{Christian2006}
{Christian}, D.~J., {Pollacco}, D.~L., {Skillen}, I., {et~al.} 2006, \mnras,
  372, 1117

\bibitem[{{Claret}(1998)}]{Claret1998}
{Claret}, A. 1998, \aaps, 131, 395

\bibitem[{{Claret} \& {Bloemen}(2011)}]{Claret2011}
{Claret}, A. \& {Bloemen}, S. 2011, \aap, 529, A75

\bibitem[{{Collier Cameron} {et~al.}(2010){Collier Cameron}, {Guenther},
  {Smalley}, {McDonald}, {Hebb}, {Andersen}, {Augusteijn}, {Barros}, {Brown},
  {Cochran}, {Endl}, {Fossey}, {Hartmann}, {Maxted}, {Pollacco}, {Skillen},
  {Telting}, {Waldmann}, \& {West}}]{CollierCameron2010}
{Collier Cameron}, A., {Guenther}, E., {Smalley}, B., {et~al.} 2010, \mnras,
  407, 507

\bibitem[{{Cowling}(1941)}]{Cowling1941}
{Cowling}, T.~G. 1941, \mnras, 101, 367

\bibitem[{{Cowling} \& {Newing}(1949)}]{Cowling1949}
{Cowling}, T.~G. \& {Newing}, R.~A. 1949, \apj, 109, 149

\bibitem[{{Daszy{\'n}ska-Daszkiewicz}
  {et~al.}(2003){Daszy{\'n}ska-Daszkiewicz}, {Dziembowski}, \&
  {Pamyatnykh}}]{DaszynskaDaszkiewicz2003}
{Daszy{\'n}ska-Daszkiewicz}, J., {Dziembowski}, W.~A., \& {Pamyatnykh}, A.~A.
  2003, \aap, 407, 999

\bibitem[{{Daszy{\'n}ska-Daszkiewicz}
  {et~al.}(2002){Daszy{\'n}ska-Daszkiewicz}, {Dziembowski}, {Pamyatnykh}, \&
  {Goupil}}]{DaszynskaDaszkiewicz2002}
{Daszy{\'n}ska-Daszkiewicz}, J., {Dziembowski}, W.~A., {Pamyatnykh}, A.~A., \&
  {Goupil}, M.-J. 2002, \aap, 392, 151

\bibitem[{{de Mooij} {et~al.}(2013){de Mooij}, {Brogi}, {de Kok}, {Snellen},
  {Kenworthy}, \& {Karjalainen}}]{DeMooij2013}
{de Mooij}, E.~J.~W., {Brogi}, M., {de Kok}, R.~J., {et~al.} 2013, \aap, 550,
  A54

\bibitem[{{Deming} {et~al.}(2012){Deming}, {Fraine}, {Sada}, {Madhusudhan},
  {Knutson}, {Harrington}, {Blecic}, {Nymeyer}, {Smith}, \&
  {Jackson}}]{Deming2012}
{Deming}, D., {Fraine}, J.~D., {Sada}, P.~V., {et~al.} 2012, \apj, 754, 106

\bibitem[{{Deupree}(2011)}]{Deupree2011}
{Deupree}, R.~G. 2011, \apj, 742, 9

\bibitem[{{Deupree} {et~al.}(2012){Deupree}, {Casta{\~n}eda}, {Pe{\~n}a}, \&
  {Short}}]{Deupree2012}
{Deupree}, R.~G., {Casta{\~n}eda}, D., {Pe{\~n}a}, F., \& {Short}, C.~I. 2012,
  \apj, 753, 20

\bibitem[{{Dupret} {et~al.}(2003){Dupret}, {De Ridder}, {De Cat}, {Aerts},
  {Scuflaire}, {Noels}, \& {Thoul}}]{Dupret2003}
{Dupret}, M.-A., {De Ridder}, J., {De Cat}, P., {et~al.} 2003, \aap, 398, 677

\bibitem[{{Eastman} {et~al.}(2010){Eastman}, {Siverd}, \&
  {Gaudi}}]{Eastman2010}
{Eastman}, J., {Siverd}, R., \& {Gaudi}, B.~S. 2010, \pasp, 122, 935

\bibitem[{{Eggen}(1956)}]{Eggen1956}
{Eggen}, O.~J. 1956, \pasp, 68, 238

\bibitem[{{Fitch}(1981)}]{Fitch1981}
{Fitch}, W.~S. 1981, \apj, 249, 218

\bibitem[{{Gillon} {et~al.}(2009){Gillon}, {Smalley}, {Hebb}, {Anderson},
  {Triaud}, {Hellier}, {Maxted}, {Queloz}, \& {Wilson}}]{Gillon2009}
{Gillon}, M., {Smalley}, B., {Hebb}, L., {et~al.} 2009, \aap, 496, 259

\bibitem[{{Goupil} {et~al.}(2000){Goupil}, {Dziembowski}, {Pamyatnykh}, \&
  {Talon}}]{Goupil2000}
{Goupil}, M.-J., {Dziembowski}, W.~A., {Pamyatnykh}, A.~A., \& {Talon}, S.
  2000, in Astronomical Society of the Pacific Conference Series, Vol. 210,
  Delta Scuti and Related Stars, ed. M.~{Breger} \& M.~{Montgomery}, 267

\bibitem[{{Gupta}(1978)}]{Gupta1978}
{Gupta}, S.~K. 1978, \apss, 59, 85

\bibitem[{{Hambleton} {et~al.}(2013){Hambleton}, {Kurtz}, {Pr{\v s}a}, {Guzik},
  {Pavlovski}, {Bloemen}, {Southworth}, {Conroy}, {Littlefair}, \&
  {Fuller}}]{Hambleton2013}
{Hambleton}, K.~M., {Kurtz}, D.~W., {Pr{\v s}a}, A., {et~al.} 2013, \mnras,
  434, 925

\bibitem[{{Herrero} {et~al.}(2011){Herrero}, {Morales}, {Ribas}, \&
  {Naves}}]{Herrero2011}
{Herrero}, E., {Morales}, J.~C., {Ribas}, I., \& {Naves}, R. 2011, \aap, 526,
  L10

\bibitem[{{Jones} {et~al.}(2001){Jones}, {Oliphant}, {Peterson},
  {et~al.}}]{Jones2001}
{Jones}, E., {Oliphant}, T., {Peterson}, P., {et~al.} 2001, {SciPy}: Open
  source scientific tools for {Python}, \url{http://www.scipy.org}

\bibitem[{{King}(1991)}]{King1991}
{King}, J.~R. 1991, Information Bulletin on Variable Stars, 3562, 1

\bibitem[{{Kov{\'a}cs} {et~al.}(2013){Kov{\'a}cs}, {Kov{\'a}cs}, {Hartman},
  {Bakos}, {Bieryla}, {Latham}, {Noyes}, {Reg{\'a}ly}, \&
  {Esquerdo}}]{Kovacs2013}
{Kov{\'a}cs}, G., {Kov{\'a}cs}, T., {Hartman}, J.~D., {et~al.} 2013, \aap, 553,
  A44

\bibitem[{{Lenz} \& {Breger}(2005)}]{LenzBreger2005}
{Lenz}, P. \& {Breger}, M. 2005, Communications in Asteroseismology, 146, 53

\bibitem[{{L{\'o}pez de Coca} {et~al.}(1990){L{\'o}pez de Coca}, {Rolland},
  {Garrido}, \& {Rodriguez}}]{LopezdeCoca1990}
{L{\'o}pez de Coca}, P., {Rolland}, A., {Garrido}, R., \& {Rodriguez}, E. 1990,
  \apss, 169, 211

\bibitem[{{Maeder}(2009)}]{Maeder2009}
{Maeder}, A. 2009, {Physics, Formation and Evolution of Rotating Stars}

\bibitem[{{Mandel} \& {Agol}(2002)}]{MAndelAgol2002}
{Mandel}, K. \& {Agol}, E. 2002, \apjl, 580, L171

\bibitem[{{Mathias} {et~al.}(1997){Mathias}, {Gillet}, {Aerts}, \&
  {Breitfellner}}]{Mathias1997}
{Mathias}, P., {Gillet}, D., {Aerts}, C., \& {Breitfellner}, M.~G. 1997, \aap,
  327, 1077

\bibitem[{{Milligan} \& {Carson}(1992)}]{Milligan1992}
{Milligan}, H. \& {Carson}, T.~R. 1992, \apss, 189, 181

\bibitem[{{Montgomery} \& {O'Donoghue}(1999)}]{Montgomery1999}
{Montgomery}, M.~H. \& {O'Donoghue}, D. 1999, Delta Scuti Star Newsletter, 13,
  28

\bibitem[{{Moya} {et~al.}(2011){Moya}, {Bouy}, {Marchis}, {Vicente}, \&
  {Barrado}}]{Moya2011}
{Moya}, A., {Bouy}, H., {Marchis}, F., {Vicente}, B., \& {Barrado}, D. 2011,
  \aap, 535, A110

\bibitem[{{Murphy} {et~al.}(2012){Murphy}, {Grigahc{\`e}ne}, {Niemczura},
  {Kurtz}, \& {Uytterhoeven}}]{Murphy2012}
{Murphy}, S.~J., {Grigahc{\`e}ne}, A., {Niemczura}, E., {Kurtz}, D.~W., \&
  {Uytterhoeven}, K. 2012, \mnras, 427, 1418

\bibitem[{{Patil} {et~al.}(2010){Patil}, {Huard}, \& {Fonnesbeck}}]{Patil2010}
{Patil}, A., {Huard}, D., \& {Fonnesbeck}, C.~J. 2010, Journal of Statistical
  Software, 35, 1

\bibitem[{{Petersen} \& {Hog}(1998)}]{Petersen1998}
{Petersen}, J.~O. \& {Hog}, E. 1998, \aap, 331, 989

\bibitem[{{Pollacco} {et~al.}(2006){Pollacco}, {Skillen}, {Collier Cameron},
  {Christian}, {Hellier}, {Irwin}, {Lister}, {Street}, {West}, {Anderson},
  {Clarkson}, {Deeg}, {Enoch}, {Evans}, {Fitzsimmons}, {Haswell}, {Hodgkin},
  {Horne}, {Kane}, {Keenan}, {Maxted}, {Norton}, {Osborne}, {Parley}, {Ryans},
  {Smalley}, {Wheatley}, \& {Wilson}}]{Pollaco2006}
{Pollacco}, D.~L., {Skillen}, I., {Collier Cameron}, A., {et~al.} 2006, \pasp,
  118, 1407

\bibitem[{{Pont} {et~al.}(2006){Pont}, {Zucker}, \& {Queloz}}]{Pont2006}
{Pont}, F., {Zucker}, S., \& {Queloz}, D. 2006, \mnras, 373, 231

\bibitem[{{Savonije} \& {Witte}(2002)}]{Savonije2002}
{Savonije}, G.~J. \& {Witte}, M.~G. 2002, \aap, 386, 211

\bibitem[{{Smith} {et~al.}(2011){Smith}, {Anderson}, {Skillen}, {Collier
  Cameron}, \& {Smalley}}]{Smith2011}
{Smith}, A.~M.~S., {Anderson}, D.~R., {Skillen}, I., {Collier Cameron}, A., \&
  {Smalley}, B. 2011, \mnras, 416, 2096

\bibitem[{{Southworth} {et~al.}(2009){Southworth}, {Hinse}, {J{\o}rgensen},
  {Dominik}, {Ricci}, {Burgdorf}, {Hornstrup}, {Wheatley}, {Anguita}, {Bozza},
  {Novati}, {Harps{\o}e}, {Kj{\ae}rgaard}, {Liebig}, {Mancini}, {Masi},
  {Mathiasen}, {Rahvar}, {Scarpetta}, {Snodgrass}, {Surdej}, {Th{\"o}ne}, \&
  {Zub}}]{Southworth2009}
{Southworth}, J., {Hinse}, T.~C., {J{\o}rgensen}, U.~G., {et~al.} 2009, \mnras,
  396, 1023

\bibitem[{{Southworth} {et~al.}(2011){Southworth}, {Zima}, {Aerts}, {Bruntt},
  {Lehmann}, {Kim}, {Kurtz}, {Pavlovski}, {Pr{\v s}a}, {Smalley}, {Gilliland},
  {Christensen-Dalsgaard}, {Kawaler}, {Kjeldsen}, {Cote}, {Tenenbaum}, \&
  {Twicken}}]{Southworth2011}
{Southworth}, J., {Zima}, W., {Aerts}, C., {et~al.} 2011, \mnras, 414, 2413

\bibitem[{{Stellingwerf}(1979)}]{Stellingwerf1979}
{Stellingwerf}, R.~F. 1979, \apj, 227, 935

\bibitem[{{Strassmeier} {et~al.}(2010){Strassmeier}, {Granzer}, {Weber},
  {Woche}, {Popow}, {J{\"a}rvinen}, {Bartus}, {Bauer}, {Dionies}, {Fechner},
  {Bittner}, \& {Paschke}}]{STELLA2010}
{Strassmeier}, K.~G., {Granzer}, T., {Weber}, M., {et~al.} 2010, Advances in
  Astronomy, 2010

\bibitem[{{Uytterhoeven} {et~al.}(2011){Uytterhoeven}, {Moya},
  {Grigahc{\`e}ne}, {Guzik}, {Guti{\'e}rrez-Soto}, {Smalley}, {Handler},
  {Balona}, {Niemczura}, {Fox Machado}, {Benatti}, {Chapellier}, {Tkachenko},
  {Szab{\'o}}, {Su{\'a}rez}, {Ripepi}, {Pascual}, {Mathias},
  {Mart{\'{\i}}n-Ru{\'{\i}}z}, {Lehmann}, {Jackiewicz}, {Hekker},
  {Gruberbauer}, {Garc{\'{\i}}a}, {Dumusque}, {D{\'{\i}}az-Fraile}, {Bradley},
  {Antoci}, {Roth}, {Leroy}, {Murphy}, {De Cat}, {Cuypers}, {Kjeldsen},
  {Christensen-Dalsgaard}, {Breger}, {Pigulski}, {Kiss}, {Still}, {Thompson},
  \& {van Cleve}}]{Uytterhoeven2011}
{Uytterhoeven}, K., {Moya}, A., {Grigahc{\`e}ne}, A., {et~al.} 2011, \aap, 534,
  A125

\bibitem[{{von Zeipel}(1924)}]{vonZeipel1924}
{von Zeipel}, H. 1924, \mnras, 84, 665

\bibitem[{{Watson}(1988)}]{Watson1988}
{Watson}, R.~D. 1988, \apss, 140, 255

\bibitem[{{Willems}(2003)}]{Willems2003}
{Willems}, B. 2003, \mnras, 346, 968

\bibitem[{{Zima}(2008)}]{Zima2008}
{Zima}, W. 2008, Communications in Asteroseismology, 157, 387

\end{thebibliography}

\end{document}